\documentclass[a4paper,usedcolumn,usenatbib]{mnras}

\usepackage[utf8]{inputenc}
\usepackage{ae,aecompl}

\usepackage{booktabs}
\usepackage[detect-all]{siunitx}

\usepackage[labelfont=bf,small]{caption}
\DeclareCaptionFormat{A1}{#1. \small{Dependence of the critical surface density for lensing on the lens redshift, computed for a source redshift of $z_s = 2$ and cosmological parameters from the \emph{Planck} 2015 results \citep{planck}. The dotted gray line marks the minimum at $z \sim 0.5$.}\\}

\usepackage{etoolbox}
\makeatletter
\patchcmd\@combinedblfloats{\box\@outputbox}{\unvbox\@outputbox}{}{%
  \errmessage{\noexpand\@combinedblfloats could not be patched}%
}
\makeatother

\usepackage{graphicx}	% Including figure files
\usepackage{amsmath}	% Advanced maths commands
\usepackage{amssymb}	% Extra maths symbols

\usepackage{mathtools}

\newcommand\eg{{\it e.g.}\ }
\newcommand\ie{{\it i.e.}\ }
\newcommand\cf{{\it cf.}\ }
\newcommand{\LCDM}{$\Lambda$CDM }

\newcommand{\hl}{\textbf}
\renewcommand{\hl}{}

\title[Efficient detection of strongly-lensing groups and clusters]{EasyCritics I: Efficient detection of strongly-lensing galaxy groups and clusters in wide-field surveys}

\author[S. Stapelberg et al.]{
Sebastian Stapelberg,$^{1}$\thanks{E-mail: stapelberg[at]stud.uni-heidelberg.de}
Mauricio Carrasco,$^{1}$
Matteo Maturi,$^{1}$
\\
$^{1}$Zentrum für Astronomie der Universität Heidelberg, Institut für Theoretische Astrophysik, Philosophenweg 12, 69120 Heidelberg, Germany.}

\date{}%\date{Accepted XXX. Received YYY; in original form ZZZ}

\pubyear{2018}

\begin{document}
\label{firstpage}
\pagerange{\pageref{firstpage}--\pageref{lastpage}}
\maketitle

\begin{abstract}
We present \emph{EasyCritics}, an algorithm to detect strongly-lensing groups and clusters in wide-field surveys without relying on a direct recognition of arcs.
\emph{EasyCritics} assumes that light traces mass in order to predict the most likely locations of critical curves from the observed fluxes of luminous red early-type galaxies in the line of sight. The positions, redshifts and fluxes of these galaxies constrain the idealized gravitational lensing potential as a function of source redshift up to five free parameters, which are calibrated on few known lenses.
From the lensing potential, \emph{EasyCritics} derives the critical curves for a given, representative source redshift.
The code is highly parallelized, uses fast Fourier methods and, optionally, GPU acceleration in order to process large datasets efficiently. The search of a $\smash{1 \, \mathrm{deg}^2}$ field of view requires less than 1 minute on a modern quad-core CPU, when using a pixel resolution of $0.25''/\mathrm{px}$.
In this first part of a paper series on \emph{EasyCritics}, we describe the main underlying concepts and present a first demonstration on data from the \emph{Canada-France-Hawaii-Telescope Lensing Survey}. We show that \emph{EasyCritics} is able to identify known group- and cluster-scale lenses, including a cluster with two giant arc candidates that were previously missed by automated arc detectors. 
\end{abstract}


%%%%%%%%%%%%%%%%%%%%%%%%%%%%%%%%%%%%%%%%%%%%%%%%%%

%%%%%%%%%%%%%%%%% BODY OF PAPER %%%%%%%%%%%%%%%%%%

\section{Introduction} \label{sec:intro}

Gravitational lensing by galaxy clusters has become a powerful tool for studying the dark universe. Strong and weak lensing signatures provide the most robust observational constraints on the projected mass distribution in clusters \citep[\eg][]{cacciato2006, Limousin07, Merten2009}.
The cluster mass function and profiles are sensitive to the late time evolution of cosmic structures and thus offer insights into fundamental questions of cosmology, such as the origin of the cosmic acceleration \mbox{\citep{Voit2005, Allen11}}.

In the dense core regions of massive groups and clusters, lensing unfolds its strong regime, where multiple images of a background source can be formed. For extended sources, these images are highly distorted and magnified, which causes them to appear in the form of luminous arcs. The analysis of arcs and multiple images allows to probe the inner cluster structure with minimal biases \citep[\eg][]{meneghetti2013}, to measure the Hubble constant \citep[\eg][]{vega2017,grillo18} and to test cosmological models \mbox{with giant arc statistics \citep[\eg][]{Bartelmann98}.}

Intriguingly, the statistics of giant arcs are observed to be in conflict with theoretical expectations. For nearly two decades, \LCDM cosmological simulations have underestimated both the abundance \citep{Bartelmann98, gladders2003, Li06, Fedeli2008EDE2, Bayliss12, Horesh2011} and the mean radii \citep[\eg][]{BroadhurstBarkana2008, Zitrin2011d} of giant arcs by differing amounts.
Although considerable progress has been made in refining simulations with improved lens and source models, the precise levels and origins of these discrepancies are still not well understood \citep{meneghetti2013, boldrin2016}.
One of the major reasons for this uncertain picture are the limited sizes of available datasets, which so far appear insufficient for a solid comparison of theory with observations.

At present, the largest homogeneous samples of \mbox{giant} arcs comprise a few hundred exemplars only.
A significant rise in the rate of detections, however, is expected with the next generation of wide-field surveys, carried out with the \emph{ESA Euclid} space mission \mbox{\citep{euclid1}} and the \emph{Large Synoptic Survey Telescope} \mbox{\citep{lsst}}. The latter promise order-of-magnitude increases in the \mbox{number} of \mbox{observed} arcs \citep{boldrin2012}, opening exciting prospects for the use of giant arcs as a competitive cosmological probe.

However, the reliable identification of strong lensing events within large image material remains a significant challenge.
Recent blind searches have focused on crowdsourced visual inspection \citep[\eg][]{sw1} and automated arc-finding algorithms \cite[\eg][]{Lenzen04,Horesh05,Alard06,seidel2007,more2012,xu2016}. 
Although many new arc candidates were discovered, especially around single galaxies, both approaches face some limitations.
Visual inspection generally lacks an objective definition of arcs. Moreover, it does not \mbox{allow} for a straightforward repetition of the analysis with new selection criteria; and, in addition, it is highly time consuming. In order to remain feasible, the efforts of increasingly large communities of trained volunteers are required {\citep[\eg][]{sw1,sw2}}.
Automated arc detectors, on the other hand, mostly suffer from a high contamination with spurious detections -- at least in the case of searches on group and cluster scales, where the features of interest can have complex morphologies. As a result, a manual validation of up to thousands of candidates per square degree is required nonetheless \citep{Maturi14}.

In order to alleviate the human bottleneck in lens detection, increasing interest has lately been devoted to machine learning methods \citep[\eg][]{Bom2017, jacobs17}. 
Nonetheless, there are good reasons to investigate alternative strategies for lens detection, as well.
Arc-based searches are naturally affected by ambiguities due to the inherent similarity of arcs with several other, commonly observed features\footnote{Especially in the case of arcs with small length-to-width ratios and images blurred by atmospheric turbulence.}. The secure identification of arcs thus requires a subsequent spectroscopic redshift measurement. For this, the candidate samples need to be sufficiently pure to optimize the follow-up efficiency. An effective decontamination of candidate samples can only be achieved at high costs, involving the use of increasingly complex -- and less physically understood -- search patterns or the exhaustion of time and human resources. Complementary to the existing approaches, however, a pre-selection of targets for arc searches, based on additional, independent indicators of the presence of strongly-lensing structures, may have the potential to remove a significant part of the aforementioned ambiguities with minimal effort and based on simple physical arguments.

Here, we present a method that identifies group- and cluster-scale strong lenses by their own optical characteristics rather than by their effect on background images.
We introduce the \emph{EasyCritics} algorithm, which predicts the locations of critical curves based on the spatial distribution of early-type luminous red galaxies (LRGs) observed in the line of sight.
Under the assumption that this distribution follows the underlying matter field, \emph{EasyCritics} develops an idealized model of the line-of-sight mass density distribution, which requires only five free parameters and is used to predict the lensing potential.
For a given source redshift, \emph{EasyCritics} then produces maps and catalogs of the resulting critical curves, 
\pagebreak
which can be used to pre-select targets for a subsequent inspection with conventional methods.
The pre-selection with \emph{EasyCritics} promises (1) a significant increase in efficiency compared to visual inspection methods; (2) a significant increase in purity compared to existing automated methods; (3) a quantitative criterion for rating the likelihood of detections; and (4) a first estimate of the strong lensing properties, such as the sizes and orientations of the critical curves.

The underlying assumption that light traces mass (LTM) is well-tested over a wide range of scales \citep[\eg][]{bahcall2014, zitrin2015}. Early-type LRGs, in particular, have been recognized as distinguished probes of the underlying density field \citep[\eg][]{Gavazzi04, zheng09, Zitrin12MACS0329a, wong2013}. 
In the context of lens modelling, the LTM assumption is routinely used for the reconstruction and prediction of multiple images in massive clusters \cite[\eg][]{Zitrin09, Zitrin2011c, LimousinOnA1689, Limousin2012_M0717, Richard2014, jauzac2015, ishigaki2015, kawamata2016, caminha2017}.
So far, however, the LTM-based analysis of lensing has primarily focused on fitting the properties of individual, known lenses.
In contrast, \emph{EasyCritics} blindly models the foreground structures over an entire line of sight and throughout large celestial regions -- not only for predefined, isolated galaxy clusters. The galaxies to be used are determined automatically according to their luminosity function and for different redshift bins. In doing so, \emph{EasyCritics} combines the lens modelling approach described by \cite{Zitrin09} and the arc-free analysis by \cite{Zitrin12MACS0329a} with the LRG-based lens detection by \cite{wong2013} and \mbox{\cite{ammons2014}}. Due to the much larger datasets at hand, the mathematical approach and numerical implementation was changed in order to achieve higher computational performances.

This is the first of a series of papers devoted to \mbox{\emph{EasyCritics}}. In this work, we describe the underlying concepts and methods; and demonstrate their feasibility by showing a few preliminary examples of the application of \emph{EasyCritics} to data from the \emph{Canada-France-Hawaii-Telescope Lensing survey} (\emph{CFHTLenS,} \citealt{Heymans12, Erben13}). In two forthcoming papers, we will furthermore provide a full statistical description of the purity and completeness, the positive detections and their lensing properties; and a comparison of the latter with other studies -- using both observational \citep[in prep.]{2nd_paper} and simulated data.

The paper is organized as follows: Section~\ref{lrg_def} summarizes the input observables and pre-processing steps. Section~\ref{mass_modelling} gives an outline of the LTM model, as we propose it, and Section~\ref{lensing} proceeds with our description of lensing.
In Section~\ref{implementation}, we then discuss details of the numerical implementation, before we conclude this work with the application examples in Section~\ref{demo}.

\newpage

\section{Observables and galaxy selection} \label{lrg_def} 

\emph{EasyCritics} takes as input a galaxy catalog, containing the coordinates $\boldsymbol{\theta}$, photometric or spectroscopic redshifts $z$ and red\footnote{{\it E.g.}\ the \emph{CFHTLenS} -- $r'$ or \emph{CFHTLenS} -- $i'$ band.}-band fluxes $F$ of all early-type\footnote{Selected for example according to their spectral energy distribution.} (E/S0) galaxies observed.
The fluxes are converted to intrinsic luminosities $L$ using the redshift information and the luminosity distance for a given background cosmological model.

\begin{figure*}
\begin{center}
\hfill
\raisebox{-0.5\height}{\includegraphics[width=1.24\columnwidth,clip,trim=1mm 0mm 0mm 0mm]{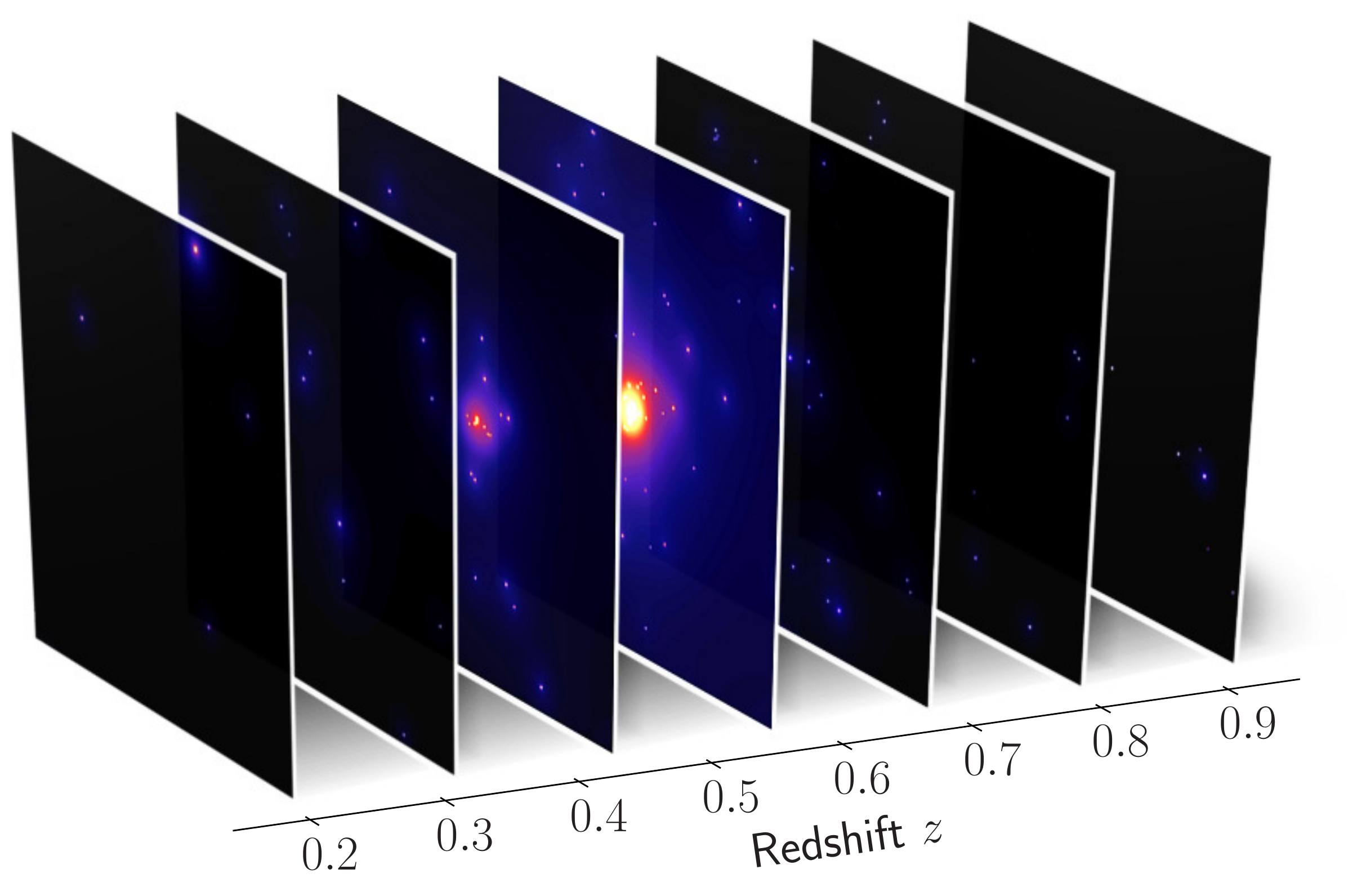}}  \hfill
\raisebox{-0.5\height}{\includegraphics[width=0.8\columnwidth,clip,trim=0mm 0mm 0mm 0mm]{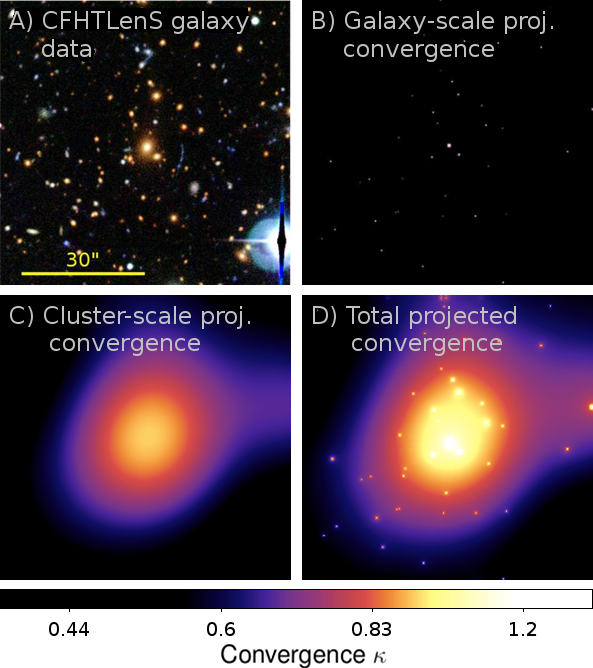}}
\caption{\label{fig:illustration_binning}
Blind identification and modelling of overdense line-of-sight structures based on LRGs, illustrated for a region centered on a cluster-scale detection. \emph{Left:} Surface density distributions $\smash{\Sigma^{(k)}}$ of the various sheets distributed along the line of sight, in arbitrary units. \emph{Right:} CFHTLenS $i' r' g'$-composite image (A) and line-of-sight projected maps of the predicted convergence at different levels of the modelling procedure (B-D). Shown are B) the projected convergence $\smash{\bar \kappa_{\mathrm{gal}}}$ of the galaxies only, C) the resulting projected convergence $\smash{\bar \kappa_{\mathrm{clus}}}$ of cluster-scale matter and D) the total projected convergence $\smash{\bar \kappa = \bar \kappa_{\mathrm{clus}} + \bar \kappa_{\mathrm{gal}}}$.
}
\end{center}
\end{figure*}

\emph{EasyCritics} selects all galaxies from a user-defined redshift interval. In this study, we restrict ourselves to the range:
\begin{align}
        0.2 \lesssim z \lesssim 0.9, \label{zl_bounds}
\end{align}
which is motivated by the results of previous studies of the redshift distribution of strong lenses with large arcs \citep[\eg][]{Bayliss11b, Bayliss12, carrasco17}.
If necessary, the range can be restricted further, so as to ensure that the characteristic magnitudes remain sufficiently below the respective limiting magnitudes and that the photometric redshift uncertainties are reliable.

The galaxy redshifts are binned, with the size of each bin set to twice the median photometric measurement uncertainty $\Delta z$ of the given input data. The position of each bin is then set to the average redshift of the galaxies it contains.
For a future implementation, we intend to take into account the whole redshift probability density distribution by accordingly distributing the contribution of each galaxy over multiple bins.

For each redshift bin, we select only the brightest galaxies with red-band absolute magnitudes below
\begin{align}
        M \lesssim M_{\star} + 2.
\end{align}
Here, $M_{\star}$ denotes the characteristic magnitude that we obtain by fitting the Schechter luminosity function \citep{Schechter76}.
This selection removes galaxies with large photometric uncertainties and would yield insignificant or even spurious contributions to the mass estimate.

%%%%%%%% TODO GOOD - DONT CHANGE UP TO HERE %%%%%%%%%%%%%

\section{Mass model} \label{mass_modelling}

\emph{EasyCritics} uses the distribution of light from LRGs for a blind, automated identification and modelling of strongly-lensing structures.
These structures are distributed over a wide range in redshift and their strong lensing signatures may be perturbed significantly by the foreground interlopers present within the line of sight \citep[\eg][]{Wambsganss05, King_Corless07, PuchweinHilbert2009, wong2013, Bayliss2014}. For this reason, we include all galaxies, selected as discussed in Section~\ref{lrg_def}, within the considered light cone. In this section, we discuss our mass model based on the LTM approach.

\subsection{Stacked mass sheets} \label{stack}

We introduce mass sheets $\smash{\Sigma^{(k)}}$ \mbox{$(k \in \mathbb N)$} at all redshift bins $\smash{z^{(k)}}$ (\cf Fig.~\ref{fig:illustration_binning}, left).
The surface density distribution on each sheet is modelled based on the LRGs belonging to the respective bin.
\hl{
In our derivation, we assume that there is only one dominant group- or cluster-scale lens within the line of sight, while the remaining structures add weak and uncorrelated local perturbations only. 
The angular separation of the lenses at different redshifts is assumed to be large enough for their deflections to be effectively independent. 
This assumption allows to neglect the lens-lens coupling between the sheets and is justified by the low probability of having chance alignments of large objects, such as the groups and clusters we are interested in \citep[\eg][and references therein]{schneider2014}. However, in the rare situation that such alignments are encountered, \emph{EasyCritics} would detect the lenses anyway unless their ability to form a critical curve depends on the higher-order interactions.
}

In analogy to cosmological weak gravitational lensing, we introduce an effective convergence $\bar \kappa$ as a weighted projection of the surface density along the line of sight:
\begin{align}
	\bar \kappa \equiv \sum_{k} \kappa^{(k)}, \label{bar_def}
\end{align}
where the contribution by each sheet is given by:
\begin{align}
        \kappa^{(k)} \equiv \frac{\Sigma^{(k)}}{\Sigma_{\mathrm{crit}}(z^{(k)}) }, \label{kappa_k}
\end{align}
and the critical surface density $\Sigma_{\mathrm{crit}}$ is generalized by: 
\begin{align}
        \Sigma_{\mathrm{crit}}(z^{(k)}) \equiv \frac{c^2}{4 \pi G} \frac{D_s}{D_l^{(k)} \, D^{(k)}_{ls}}.
\end{align}
As usual, $z_s$ denotes the source redshift and the symbols $\smash{D_s \equiv D(0, z_s)}$, $\smash{D_l^{(k)} \equiv D(0, z^{(k)})}$ and $\smash{D^{(k)}_{ls} \equiv D(z^{(k)}, z_s)}$ abbreviate the angular-diameter distances $D(z_1, z_2)$ between two redshifts $z_1$ and $z_2$.

\subsection{Surface density estimation}

On each sheet, the total surface density distribution is modelled as a smooth cluster-scale component $\smash{\Sigma^{(k)}_{\mathrm{clus}}}$ and a superposition $\smash{\Sigma^{(k)}_{\mathrm{gal}}}$ of embedded galaxy-scale subhalos:
\begin{align}
        \Sigma^{(k)} = \Sigma^{(k)}_{\mathrm{clus}} + \Sigma^{(k)}_{\mathrm{gal}}. \label{sigma_total}
\end{align}
Here, both components refer to total surface densities, \ie they are not further distinguished into dark, gaseous and stellar contributions. Thus, small offsets or misalignments between the luminous and dark matter halos are neglected.

This two-component modelling approach is motivated by a number of previous strong-lensing analyses of massive clusters, 
where its accuracy and efficiency has been shown \citep[\eg][]{Broadhurst05,Smith2001,Zitrin09}.
The modelling of mass down to galaxy scales allows to resolve smaller granularity in the lens matter. Although most member galaxies in groups and clusters are expected to have only little influence on the formation of arcs \citep{KassiolaKovnerFort92, Kneib96, meneghetti2000}, tidal field contributions by individual bright members, such as cD-galaxies, can perturb the strong lensing cross section significantly \mbox{\citep{Meneghetti2003}}.

The components $\smash{\Sigma^{(k)}_{\mathrm{clus}}}$ and $\smash{\Sigma^{(k)}_{\mathrm{gal}}}$ are related to the LRG observables via the blind LTM modelling approach described and explained in detail throughout the subsections \ref{ml_conv1} and \ref{ml_conv2}.
The LTM modelling approach is ideally suited for the purpose of a first estimation of strong lensing properties, since it provides a highly flexible mass model at a minimum number of required model parameters -- allowing to recover the often important asymmetries and substructures in the cluster matter distribution \citep[\eg][]{BSW95,Meneghetti07}.
In particular, the LTM model introduced here has only five free parameters, which are calibrated by fitting critical curves to the locations of the arcs of few known lenses.

\subsection{Galaxy-scale subhalos} \label{ml_conv1}

Galaxy-scale subhalos are assigned to all LRGs that are selected from the line of sight according to the criteria outlined in Section~\ref{lrg_def}, including the majority of galaxies that may not be bound to a strongly-lensing group or cluster. 
Each galaxy-scale subhalo is modelled as an axially symmetric density contribution, since the influence of intrinsic ellipticities on the tidal shear perturbations is negligible. 

We assign to each subhalo a power-law profile with the same free index $q > 0$ and an amplitude that scales linearly with the galaxy luminosity $L_i$ \citep{brainerd96}:
\begin{align}
        \Sigma_{\mathrm{gal},i}(\theta) = K_{\mathrm{gal}} L_i \cdot \big( D_l \theta \big)^{-q}. \label{sigma_gal}
\end{align}
Here, $K_{\mathrm{gal}}$ is a free parameter expressing the $M/L$ proportionality, $\theta \equiv \| \boldsymbol \theta \|$ refers to the angular distance from the galaxy center and $D_l$ denotes the angular-diameter distance to the respective redshift bin of the galaxy. When calibrating the parameters, we restrict the power-law index $q$ to the interval \mbox{$q \in (0, 2)$} to ensure a well-defined lensing potential on the whole domain $\smash{\mathbb R^2}$. Note that the special case of a singular isothermal sphere ($q = 1$) is enclosed by this choice. 

The surface density contribution due to all galaxies on the sheet can be obtained by superposition:
\begin{align}
        \Sigma^{(k)}_{\mathrm{gal}}(\boldsymbol \theta) = K_{\mathrm{gal}} \cdot \big( D_l^{(k)} \big)^{-q} \sum_{  \mathclap{ \smash{ \hphantom{z^{(k)}} z_i \, = \, z^{(k)} \hphantom{z_i} } }  } \; L_i \| \boldsymbol \theta - \boldsymbol \theta_i \|^{-q},
\end{align}
where the index $i$ runs over all galaxies binned to the $k$-th sheet at $\smash{z^{(k)}}$. Here, $\boldsymbol \theta_i$ and $z_i$ denote the position and redshift bin of the $i$-th galaxy, respectively.

\subsection{Cluster-scale component} \label{ml_conv2}

According to the LTM assumption, high concentrations of LRGs trace the peaks of the underlying density field, such as cluster-scale\footnote{In order to simplify the nomenclature, we will use the term 'cluster' to refer to groups as well unless an explicit distinction is made.} halos. Within these halos, the density distribution of matter can be approximated by the smoothed distribution of LRGs \citep{zitrin2015}.
Of course, not all galaxies are associated with a cluster environment. Isolated field galaxies, for instance, need to be excluded from the $\smash{\Sigma^{(k)}_{\mathrm{clus}}}$ component. We  distinguish between cluster and field regions by introducing a weight $\smash{w^{(k)}}$, defined to be $\smash{w^{(k)} = 1}$ in supercritical cluster environments and $\smash{w^{(k)} = 0}$ in voids.

Our ansatz is to find a linear operator $\smash{\hat O}$ that approximates the cluster-scale surface density distribution $\smash{\Sigma^{(k)}_{\mathrm{clus}}}$ as:
\begin{align}
	\Sigma^{(k)}_{\mathrm{clus}} =  \hat O \left( w^{(k)} \frac{ K_{\mathrm{clus}} }{K_{\mathrm{gal}} } \Sigma^{(k)}_{\mathrm{gal}}   \right), \label{clus_tracing}
\end{align}
where suitable expressions for $\smash{\hat O}$ and $\smash{w^{(k)}}$ still need to be found and where an additional free parameter $\smash{K_{\mathrm{clus}}}$ determines the $M/L$ ratio of the cluster-scale component.
The linear scaling of mass with luminosity is expected to be a good approximation on cluster scales that should improve with richness \citep[\eg][]{lin2003, lin2004, tinker2005}.

In dense cluster environments ($w^{(k)} \sim 1$), the distribution $\smash{\Sigma^{(k)}_{\mathrm{clus}}}$ of surface mass density can be well-approximated by the galaxy-scale surface density distribution $\smash{\Sigma^{(k)}_{\mathrm{gal}}}$, once the latter is appropriately smoothed. This suggests to idealize the operator $\smash{\hat O}$ by a convolution with a smoothing window. The simplest empirically motivated choice for the latter is a Gaussian \citep{zitrin2013}:
\begin{align}
S(\boldsymbol \theta) \equiv \frac 1{2 \pi \sigma^2} \mathrm{exp} \left( - \frac{\boldsymbol \theta^2}{2 \sigma^2} \right), \label{gaussian}
\end{align}
which has only one free parameter, the standard deviation $\sigma$. The latter affects the scale of halos for the component $\smash{\Sigma^{(k)}_{\mathrm{clus}}}$. 
\hl{
We neglect any explicit dependence of $\sigma$ on the cluster concentrations because the latter is to a large part taken into account by using the actual distribution of galaxies.
}

A suitable expression for the weight $\smash{w^{(k)}}$ is determined in the following. We first quantify the local environment based on the average local number density $n(\boldsymbol \theta)$ of LRGs within a local region with area $A \gg \pi \sigma^2$, where a typical value could be $A = 15'\times15'$. We introduce a free parameter $n_c$ that specifies a 'critical' number density required for overdense regions to satisfy the criterion $\smash{w^{(k)} = 1}$. 
For simplicity, a possible redshift dependence of $n_c$ is neglected. 
The weight $\smash{w^{(k)}}$ is a functional of the relative LRG abundance, $\smash{w^{(k)}[n^{(k)}(\boldsymbol \theta)/n_c]}$, which can be interpreted as a conditional probability for LRGs to trace a cluster-scale halo given a local number density $\smash{n^{(k)}}$ of these LRGs.
A simple expression for $\smash{w^{(k)}}$ would be a Heaviside step function, $\smash{\Theta(n^{(k)}-n_c)}$. However, rather than a sharp threshold, we expect a gradual transition from $\smash{w^{(k)} = 0}$ to $\smash{w^{(k)} = 1}$, which needs to be accounted for in order not to miss possibly important galaxy- and group-scale halos. We therefore adopt the following functional form,
\begin{align}
        w^{(k)} \left[n^{(k)}\right] \equiv \left\{ \begin{array}{ll} \frac { n^{(k)}}{n_c} \mathrm e^{- 5.6 \left(\frac { n^{(k)} }{n_c} -1 \right)^2} & \text{for } n^{(k)} \leq n_c \\ 1 & \text{else} \end{array}, \right.
\end{align}
which is motivated empirically to satisfy the aforementioned criteria.

In conclusion, the previous considerations suggest to describe the cluster-scale surface density distribution $\Sigma_{\mathrm{clus}}$ by an expression of the form:
\begin{align}
        \Sigma^{(k)}_{\mathrm{clus}}(\boldsymbol \theta) = K_{\mathrm{clus}} \left( S * w^{(k)} \frac{\Sigma^{(k)}_{\mathrm{gal}}}{K_{\mathrm{gal}}} \right) (\boldsymbol \theta), \label{sigma_clus}
\end{align}
where
\begin{align}
        (f * g)(\boldsymbol \theta) \equiv \int_{\mathbb R^2} f(\boldsymbol \theta - \boldsymbol \theta') \, g(\boldsymbol \theta') \, \mathrm d^2 \theta'
\end{align}
denotes the convolution between two functions $f$ and $g$ in two dimensions.

\begin{table}
        \centering
        \caption{Summary of LTM model parameters.}
        \label{table:params}
        \begin{tabular}{ll} 
                \toprule
		\toprule
                Symbol & Description\\ 
                \midrule
		$q$ & Power-law index\\
		$K_{\mathrm{gal}}$ & $\Sigma/L$ conversion constant for LRGs\\
		$K_{\mathrm{clus}}$ & $\Sigma/L$ conversion constant for galaxy clusters \\
		$\sigma$ & Standard deviation for the Gaussian smoothing\\
		$n_c$ & Critical number density\\
                \bottomrule
        \end{tabular}
\end{table}
All free parameters introduced in the previous two subsections are summarized in Table~\ref{table:params}.

\section{Strong lensing prescription} \label{lensing}

\subsection{General assumptions} \label{lensing_assumptions}

\hl{
For our physical description of lensing, we adopt an idealized version of the multiple lens-plane framework, in which we introduce multiple geometrically thin mass sheets $\smash{\Sigma^{(k)}}$ 
under the assumption that these sheets are uncorrelated and contain a dominant strongly-lensing object embedded in weakly-lensing structures, as stated before in Subsection~\ref{stack}.
The lens equation and its Jacobian are expanded to first order in the lensing potential, thus neglecting corrections due to Born's approximation or lens-lens coupling, as described in Subsection~\ref{lens_equation}.
}

Throughout all calculations, we assume a fixed source plane that can be defined by the user.
Our default choice is $z_s = 2$, a value representative for the typical giant arc population according to broadband photometric \citep[\eg][]{Bayliss12} and spectroscopic studies \citep[\eg][]{Bayliss11a, carrasco17}. For sources at $z_s = 2$, lenses are expected to be most efficient at $\smash{z^{(k)} \sim 0.5}$ (\cf Fig.~\ref{fig:crit_density}), which is consistent with our choice Eq.~\ref{zl_bounds} for the bounds on the lens redshifts $\smash{z^{(k)}}$. However, it is worth to point out that the results predicted by \emph{EasyCritics} are not very sensitive to this number, as we demonstrate in Section~\ref{demo}.

\subsection{Effective lens equation} \label{lens_equation}

\hl{
In the multiple lens-plane approximation, observed angles $\boldsymbol{\theta}$ of images are related to the true source positions $\boldsymbol \beta(\boldsymbol \theta)$ by the following lens equation \citep{Schneider92book}:
\begin{align}
	\boldsymbol{\beta}(\boldsymbol \theta) = \boldsymbol{\theta} - \sum_{k=1}^N \boldsymbol{\alpha}^{(k)}( \boldsymbol{\theta}^{(k)}). \label{lens_eq}
\end{align}
Here, $N$ is the number of lensing mass sheets and \mbox{$\boldsymbol{\alpha}^{(k)} = D_{ls}^{(k)}/D_s \, \hat{\boldsymbol{\alpha}}^{(k)}$} are the individual reduced deflection angles. As in the single lens-plane theory, the deflection angles are derived from the gradient of lensing potentials $\smash{\psi^{(k)}}$ such that $\smash{ \boldsymbol{\alpha}^{(k)} = \boldsymbol{\nabla}_{\boldsymbol \theta^{(k)}} \psi^{(k)}}$. 

In general, the lensing distortions couple in a non-linear way because of the recursive dependency of the sheetwise deflections. 
However, as discussed in Subsection~\ref{lensing_assumptions}, we assume that the lens planes are fully independent, allowing to expand Eq.~\ref{lens_eq} to first order in the lensing potential by replacing $\smash{ \boldsymbol \alpha^{(k)}(\boldsymbol \theta^{(k)}) \approx \boldsymbol \alpha^{(k)}(\boldsymbol \theta) = \boldsymbol{\nabla}_{\boldsymbol \theta} \psi^{(k)}}$. Hereafter, we will abbreviate $\boldsymbol{\nabla} \equiv \boldsymbol{\nabla}_{\boldsymbol \theta}$.
It follows that the total deflection in the lens equation~\ref{lens_eq} can be expressed as the gradient of an effective lensing potential $\smash{\bar \psi \equiv \sum_k \psi^{(k)}}$:
\begin{align}
        \boldsymbol{\beta}(\boldsymbol \theta) = \boldsymbol{\theta} - \boldsymbol{\nabla} \bar \psi(\boldsymbol{\theta}). \label{lens_eq2}
\end{align}
}
The critical curves are derived as the sets of points with a singular Jacobian $\smash{\mathbf J_{ij} \equiv \partial \beta_i / \partial \theta_j} $:
\begin{align}
        \det \mathbf J = 1 - \boldsymbol \nabla^2 \bar \psi + (\partial_1^2 \bar \psi)(\partial_2^2 \bar \psi) - (\partial_1 \partial_2 \bar \psi)^2 = 0. \label{jac_det}
\end{align}
Accordingly, the decomposition of the Jacobian yields radial ($+$) and tangential critical curves ($-$) as the individual roots 
\begin{align}
        \lambda_{\pm} = 1 - \frac{\boldsymbol \nabla^2 \bar \psi}2 \pm \sqrt{ \left[ \frac 12 (\partial_1^2 - \partial_2^2) \bar \psi \right]^2 + (\partial_1 \partial_2 \bar \psi)^2} = 0 \label{eigenvalues}
\end{align}
of the radial and tangential eigenvalues $\lambda_{\pm}$, respectively.

\subsection{The lensing potential} \label{lensing_pot}

The Poisson equation determines the effective lensing potential $\bar \psi$ for a given $\bar \kappa$ and boundary or gauge conditions. Due to the linearity of the Laplacian, we can express the Poisson equation in the following, compact form:
\begin{align}
        \boldsymbol \nabla^2 \bar \psi(\boldsymbol \theta) = 2 \bar \kappa(\boldsymbol \theta), \label{poisson}
\end{align}
where $\bar \kappa$ is the effective convergence defined in Eq.~\ref{bar_def}. A numerically convenient integral form of the Poisson equation can be derived from the superposition principle, which allows to construct solutions by means of the Green's function for the Laplacian. For a system composed of multiple, axially-symmetric halos, we can sum the single halo potentials, for which the analytic solution is straightforward.

\hl{
To maximize the numerical efficiency, we derive all lensing properties starting from the lensing potential instead of the deflection angle, in contrast to previous studies. This has the advantage of capturing all physical properties of lensing in a scalar instead of a vector field, thus (1) enabling an efficient handling of memory; (2) providing a significant reduction of the runtime by reducing the number of convolutions necessary for deriving the Jacobian; and (3) avoiding possible artifacts that may arise from the convolutions because of border effects and the cuspy profiles of galaxies. In the particular case of the power-law halo density profiles Eq.~\ref{sigma_gal}, an additional benefit of deriving all quantities from the lensing potential instead of the deflection angle is the finiteness of $\bar{\psi}$ at the galaxy centers, where the deflections become singular for $q > 1$. We now discuss in detail the analytic and numerical steps in computing the gravitational lensing potential for our LTM model.
}

\subsection{Galaxy-scale subhalo lensing potential} \label{lp1}

We start with the analytic solution of the lensing potential for a single galaxy-scale subhalo $i$ on a given sheet at angular-diameter distance $D_l$. For brevity, we temporarily drop superscripts throughout the next paragraphs. 
The most general axially-symmetric solution of Eq.~\ref{poisson} for the power-law surface density profile in Eq.~\ref{sigma_gal} is given by:
\begin{align}
	\psi_{\mathrm{gal}, i}(\theta) = \frac{2 K_{\mathrm{gal}} L_i D_l^{-q}}{\Sigma_{\mathrm{crit}} (2-q)^2}  \theta^{2-q} + C_1 \ln \theta + C_2. \quad \;\;\; \label{poisson_sol}
\end{align}
Without loss of generality and since we made the assumption $q < 2$, we may fix the gauge such that $\psi(0) = 0$, implying $C_1 = 0 = C_2$. The remaining expression can be written in the form:
\begin{align}
	\psi_{\mathrm{gal}, i}(\theta) = \tilde K_{\mathrm{gal}} L_i \theta^{2-q}, \label{psi_2}
\end{align}
where we abbreviated the constant prefactors by:
\begin{align}
        \tilde K_{\mathrm{gal}} \equiv \frac {2 D_l^{-q} K_{\mathrm{gal}} }{\Sigma_{\mathrm{crit}} (2-q)^2} .
\end{align}
As a next step, we generalize the result to $N$ galaxies on the same sheet by superposition:
\begin{align}
        \psi_{\mathrm{gal}}(\boldsymbol{\theta}) &= \sum_{k = 1}^N \psi_{\mathrm{gal}, k} (\boldsymbol \theta - \boldsymbol \theta_k) \\
				&= \tilde K_{\mathrm{gal}} \sum_{k = 1}^N L_k \| \boldsymbol \theta - \boldsymbol \theta_k \|^{2-q}, \label{superposed_psi}
\end{align}
where $\boldsymbol \theta_k$ denotes the $k$-th galaxy position.

The linear superposition of all galaxy contributions, Eq.~\ref{superposed_psi}, is formally equivalent to a convolution
\begin{align}
	\psi_{\mathrm{gal}} = \tilde K_{\mathrm{gal}} \, ( G * L ) \label{psi_gal}
\end{align}
of an \emph{effective} luminosity density $L$, defined via the Dirac delta distribution $\smash{\delta_{D}^{(2)}}$ in two dimensions:
\begin{align}
        L(\boldsymbol{\theta}) \equiv \sum_{k = 1}^N L_k \, \delta_{D}^{(2)}(\boldsymbol{\theta} - \boldsymbol{\theta}_k),
\end{align}
with the kernel:
\begin{align}
	G (\boldsymbol \theta) \equiv \| \boldsymbol \theta \| ^{2-q}. \label{fkernel1}
\end{align}

\hl{Despite the fact that $G$ has a divergent Fourier transform for the general case $q \in [0, 2)$, its fast Fourier transform, after discretizing its spatial representation over the pixel coordinates, is well-defined everywhere and enables a very efficient evaluation of Eq.~\ref{superposed_psi}.
}
The convolution technique of Eq.~\ref{psi_gal} for lens modelling in Fourier space has been applied earlier by \cite{BW94} and \cite{Puchwein05}, who gave a detailed discussion on the limits of its accuracy.

\subsection{Cluster-scale halo lensing potential} \label{lp2}

For the contribution $\psi_{\mathrm{clus}}$ of smooth cluster-scale halos to the lensing potential, we proceed in full analogy to the galaxy-scale case. First, we note that the superposition principle relates $\psi_{\mathrm{clus}}$ to $\kappa_{\mathrm{clus}}$ by the convolution:
\begin{align}
	\psi_{\mathrm{clus}} = \mathcal G * 2\kappa_{\mathrm{clus}},
\end{align}
where $\mathcal G(\boldsymbol \theta) = \ln \| \boldsymbol \theta \|/(2\pi) $ is the Green's function for the Laplacian in two dimensions. For $\kappa_{\mathrm{clus}}$, we insert Eq.~\ref{sigma_clus} and arrive at the following expression:
\begin{align}
        \psi_{\mathrm{clus}} \propto  S * \mathcal G * \left(w \, \kappa_{\mathrm{gal}} \right). \label{psi_dm_prop}
\end{align}
Since the weight $w$ is assumed to vary significantly only on scales much larger than the cluster halo scale, $d \gg \sqrt{\pi} \sigma$ (\cf Subsection~\ref{ml_conv2}), we can safely approximate Eq.~\ref{psi_dm_prop} by:
\begin{align}
        \psi_{\mathrm{clus}} &\propto w \, (S * \psi_{\mathrm{gal}}). \label{psi_dm_intermediate}
\end{align}
Inserting the previous result for $\psi_{\mathrm{gal}}$ from Eq.~\ref{psi_gal} and combining the proportionality constants and the weight as $\smash{\tilde K_{ \mathrm{clus}} \equiv w \tilde K_{\mathrm{gal}} K_{\mathrm{clus}} / K_{\mathrm{gal}}}$, we obtain:
\begin{align}
	\psi_{\mathrm{clus}} = \tilde K_{ \mathrm{clus}} \, (S * G) * L, \label{psi_dm}
\end{align}
with the brackets denoting the most efficient convolution scheme. In particular, it is important to note that the smoothed kernel 
\begin{align}
	C(\boldsymbol \theta) \equiv (S * G)(\boldsymbol \theta) \label{C_def}
\end{align}
needs to be evaluated only once and stored so that it can be used for any arbitrary dataset $L$.

As before, we explicitely derive an analytic expression for the spatial representation of the kernel $\smash C$, from which the discrete Fourier transform is computed.
This avoids the aforementioned problem of singularities in the kernel. Compared to a numerically performed Gaussian smoothing of $G$, the evaluation of the analytic result saves a\pagebreak significant amount of runtime. The convolution in Eq.~\ref{C_def} yields (\cf Appendix~\ref{App:AppendixA}):
\begin{align}
        C(\boldsymbol \theta) &= \frac 1{2\pi \sigma^2} \int_{\mathbb R^2} \exp \left(- \frac{\| \boldsymbol \theta - \boldsymbol \theta^{\prime} \|^2}{2 \sigma^2} \right) \|\boldsymbol \theta' \|^{2-q} \, \mathrm d^2 \theta' \;\; \label{G_clus} \\
        &= \frac{2^{1- \frac q2}}{\sigma^{q-2}} \, \Gamma \left(2- \frac q2 \right) {}_1 F_1 \bigg(\frac q2 - 1; 1; -\frac{\theta^2}{2 \sigma^2} \bigg), \label{final_C}
\end{align}
where ${}_1 F_1 (\alpha;\beta;z)$ is the Kummer confluent hypergeometric function of the first kind, as defined by \citep[\eg][eq. 9.210.1]{gradshteyn85}:
\begin{align}
        {}_1 F_1 (\alpha;\beta;z) \equiv \sum_{k= 0}^{\infty} \frac{(\alpha)_k}{(\beta)_k} \frac{z^k}{k!}, \quad (\beta \notin \mathbb Z_{\leq 0})
\end{align}
with the Pochhammer symbols $(x)_k \equiv \Gamma(x+k)/\Gamma(x)$.

An alternative way to arrive at the solution would be to evaluate the convolution $(S*G)$ in Fourier space using the discrete Fourier transforms of $S$ and $G$. However this does not improve the accuracy or the computational performance on the setup tested here and would be less elegant. 

\subsection{Total lensing potential}

Combining Eq.~\ref{psi_gal} with Eq.~\ref{psi_dm} and reintroducing the superscript notation for the sheet indices, we obtain the following expression for the lensing potential for a single sheet $\smash{\Sigma^{(l)}}$:
\begin{align}
        \psi^{(l)} &= \psi^{(l)}_{\mathrm{gal}} + \psi^{(l)}_{\mathrm{clus}} \\
                &= \tilde K_{\mathrm{gal}}^{(l)} \, (G * L^{(l)}) + \tilde K^{(l)}_{\mathrm{clus}}  \left( C * L^{(l)} \right). \label{psi_sum}
\end{align}
By superposing the single-sheet solutions and interchanging the order of superposition and convolutions, we can simplify this expression to two convolution integrals:
\begin{align}
        \bar \psi =  G * \sum_{l} \tilde K^{(l)}_{\mathrm{gal}} L^{(l)}   + C * \sum_{l} \tilde K^{(l)}_{\mathrm{clus}}   L^{(l)}. \label{final_res_psi}
\end{align}
Note that two is the minimum number of convolutions to be performed numerically because of the sheetwise density-dependent weight terms $\smash{w^{(l)}}$ that enter into $\smash{\tilde K^{(l)}_{\mathrm{clus}}}$.

\section{Numerical implementation} \label{implementation}

After having introduced the mathematical formalism adopted in this work, we now briefly summarize the most important details of the numerical implementation.

\subsection{Parallel computation on mesh grids} \label{parallel_comp}

The \emph{EasyCritics} code is highly optimized both for a sequential and a parallel processing on the most common CPUs. An efficient memory management ensures that \emph{EasyCritics} is optimized for large field of views covered by the current and upcoming wide-field surveys.

\begin{figure}
\centerline{
        \includegraphics[width=0.85\columnwidth, trim= 0mm 0mm 0mm 0mm,clip]{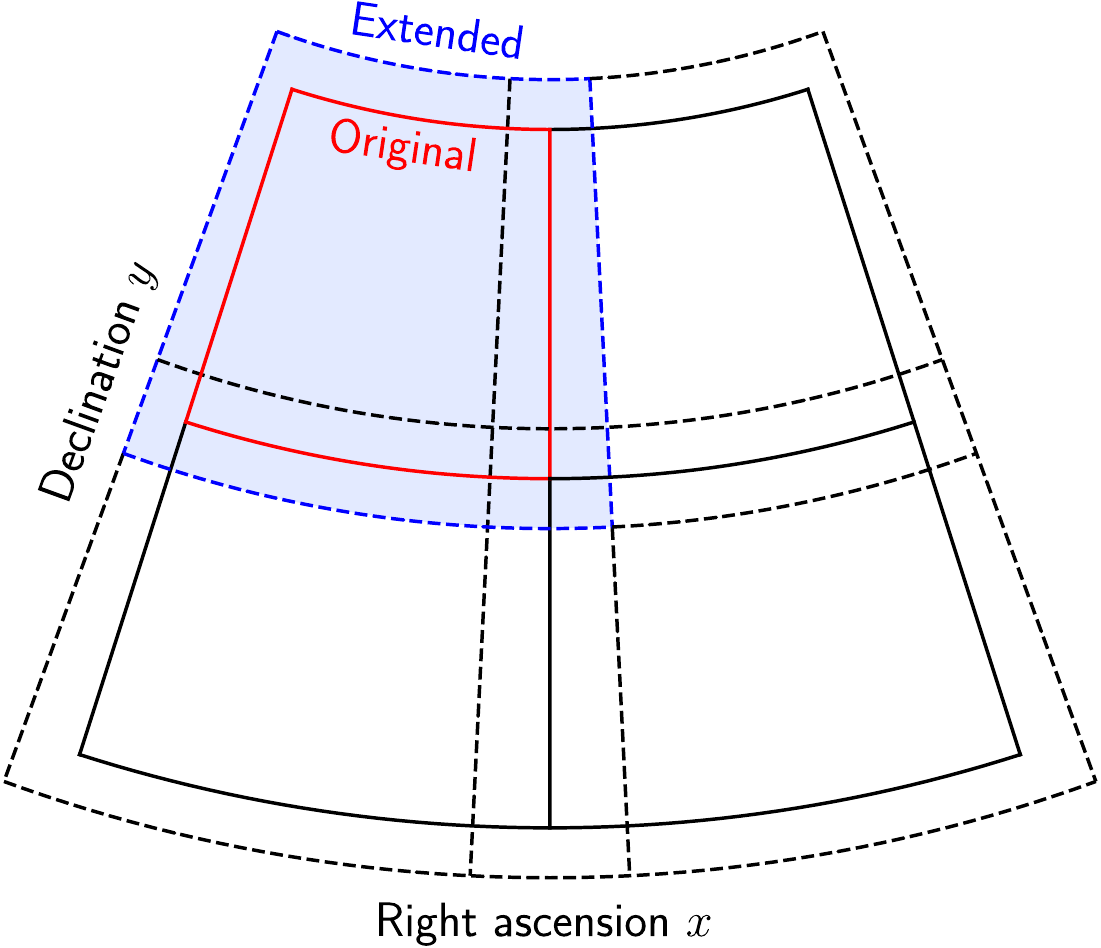}
}
\caption{Illustration of the tiled geometry in the gnomonic projection. \emph{Solid boxes:} Original tiles. \emph{Dashed boxes:} Temporarily extended tiles. For a better distinction, the first tile is colored.
\label{fig:overlapping_cells}}
\end{figure}

The total search area on the sky is divided into an array of tiles, which are processed independently and parallely by multiple threads. The number and size of these tiles, as well as the number of threads, is defined by the user.
During the computations, each tile is temporarily extended by a buffer area necessary to account for the most important gravitational contributions by adjacent matter, as illustrated in Fig.~\ref{fig:overlapping_cells}. Without such an extended area, the prediction would become increasingly inaccurate towards the tile boundaries and could lead to the loss of critical curves extending over neighboring tiles.
In addition, the temporary buffer alleviates numerical artifacts introduced by the discrete Fourier transform used to solve Eqs.~\ref{psi_gal} and \ref{psi_dm}.
The relative increase in the area of the tiles is specified by the user. A typical example setting would be an effective area of $5'\times 5'$, temporarily extended on each side by up to a factor of $3$ to ensure continuity in the lensing potential at the tile boundaries.

The computations are performed on regular mesh grids with default resolutions of $0.25''/\mathrm{px}$.
Celestial coordinates are locally mapped to the tangent plane by a gnomonic projection \citep[\eg][]{calabretta2002}.
A vector-valued grid is defined for the weighted effective luminosity maps $\smash{\tilde K^{(l)}_{\mathrm{gal}} L^{(l)}}$ and $\smash{\tilde K^{(l)}_{\mathrm{clus}} L^{(l)}}$, from which scalar-valued grids are derived via line-of-sight projection and convolved with the kernels $G$ and $C$ according to Eq.~\ref{final_res_psi}.
The confluent hypergeometric function ${}_1F_1$ is evaluated using common approximation schemes \citep[\eg][]{Muller2001,pearson2017}.

For the numerical solution of the convolutions, we use discrete Fast Fourier methods \citep{cooley1965algorithm} and exploit the discrete convolution theorem \citep{cochran67}, which allows to substitute the Fourier transform of a convolution by a product of Fourier transforms.
For the subsequent differentiation of the lensing potential, we apply suitable definitions of forward, backward and central finite differences. The angular finite differences are locally approximated by Cartesian finite differences. 

\subsection{Critical curve recognition} \label{cc_recognition}

\emph{EasyCritics} computes the Jacobian eigenvalues and the determinant based on Eq.~\ref{eigenvalues} by finite differentiation of the lensing potential. It derives a binary map of regions with a negative determinant by applying a Heaviside step function:
\begin{align}
        B(\boldsymbol \theta) = \Theta(-\det \mathbf J).
\end{align}
It then retrieves the contours of all connected components $\{ \boldsymbol \theta : B(\boldsymbol \theta) = 1 \} $ via border following \citep{suzuki1985}. As an option, radial critical curves can be removed or grouped together with the tangential lines.

\begin{figure}
\centerline{\includegraphics[width=\columnwidth, trim= 0mm 0mm 0mm 0mm,clip]{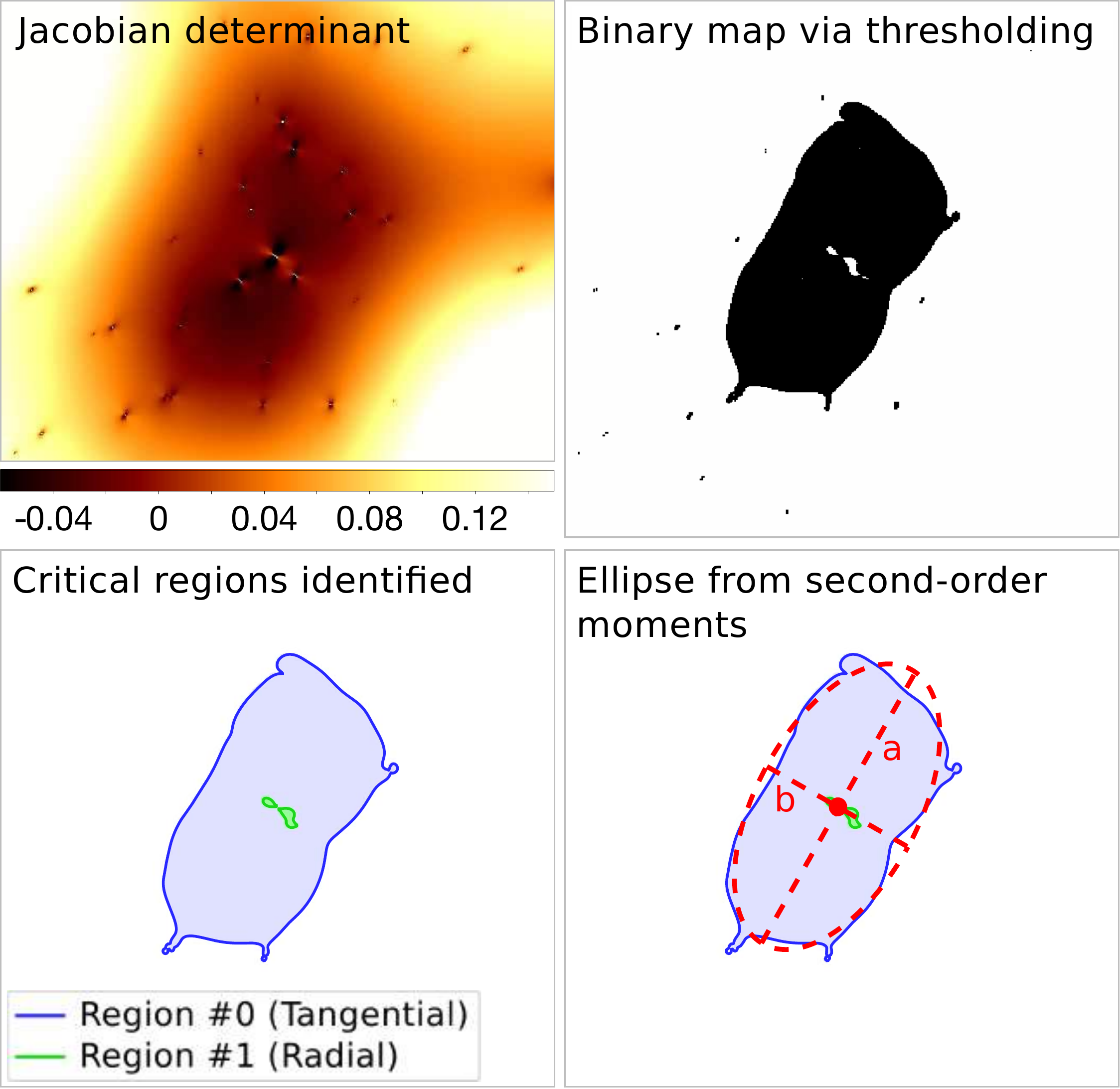}}
\caption{\label{fig:cc_rec}
        Illustration of the critical curve recognition for a region centered on a cluster-scale detection. \emph{Top left}: Jacobian determinant. \emph{Top right:} Binary map $B(\boldsymbol \theta)$ from thresholding. 
\mbox{\emph{Bottom left:}} Detected regions and their contours, displayed as filled polygons. \emph{Bottom right:} Ellipse from the moment calculation for the tangential critical curve.}
\end{figure}

For each contour, the centroid $(\bar{\theta}_1, \bar{\theta}_2)$, the orientation $\phi$ and the principal axis lengths $a,b$ are determined by an \mbox{analysis} of second-order image moments \citep{moments2, stobie86}, as defined in Appendix~\ref{appendix_moments}. The geometrical parameters determined in this way enable a selection of targets for subsequent searches based on position and size.
In order to avoid pixel artifacts due to the central cusp in our power-law description of the galaxy-scale subhalos, \emph{EasyCritics} discards all detections with an effective Einstein radius below a certain threshold, defined as the equivalent circle radius, $\theta_{\mathrm{E,eff}} \equiv \sqrt{A/\pi}$ \citep{PuchweinHilbert2009, zitrin2011}. In this study, we apply the empirically motivated value $\theta_{\mathrm{E,eff}} \sim 1.5''$.

The different steps of the recognition algorithm are illustrated for an example cluster-scale detection in Fig.~\ref{fig:cc_rec}.
For each set of critical pixels, the pixel coordinates are stored in critical curve catalogs together with the geometrical parameters $\bar{\theta}_1$, $\bar{\theta}_2$, $\phi$, $a$, $b$ and the effective Einstein radius $\theta_{\mathrm{E,eff}}$.

\subsection{Parameter calibration} \label{calib} 

The free model parameters $\smash{P = \{q, \sigma, \tilde K_{\mathrm{gal}}, \tilde K_{\mathrm{clus}}, n_c \}}$ are calibrated by fitting critical curves to the arcs of previously known lenses. 
During the calibration, we use a physically equivalent, but more convenient representation of the model parameters, $\smash{P' = \{q, \sigma, K, \mu_{\mathrm{clus}}, n_c\}}$, by introducing the two independent quantities:
\begin{align}
        K &\equiv K_{\mathrm{gal}} + K_{\mathrm{clus}}, \nonumber \\
        \mu_{\mathrm{clus}} &\equiv \frac{K_{\mathrm{clus}}}{K_{\mathrm{gal}} + K_{\mathrm{clus}}}.
\end{align}
This allows us to consider the relative amplitudes of the two mass components independently and to better cope with inherent lensing degeneracies between $q$, $\smash{K_{\mathrm{gal}}}$ and $\smash{K_{\mathrm{clus}}}$.

\hl{
Since arcs and multiple images are expected to form in the vicinity of critical curves, we use their locations and geometry to constrain the approximate positions of the critical curves \citep[\eg][]{nb96,meneghetti2007}. In a least-squares approach, we derive the free parameters $P$ by fitting the squared tangential\footnote{The method is, however, not limited to tangential eigenvalues. Radial arc data could be included as well, with each of the two eigenvalues minimized at the respective arc locations.} Jacobian eigenvalues $\lambda(P, \boldsymbol \theta_i)$ within a set of $N$ \emph{“arc pixels”} $\Theta_{\mathrm{crit}}$ that are assigned to be part of the critical curve:
\begin{align}
        \chi^2(P) = \frac 1N \sum_{ \boldsymbol \theta_i \in \Theta_{\mathrm{crit}}} \vert \lambda(P, \boldsymbol \theta_i) \vert^2. \label{objective_function}
\end{align}
This expresses the expectation that the eigenvalues vanish along the critical curve traced by the reference pixels, given a set of parameters $P$. The eigenvalue-based criterion provides a local and robust least-squares estimate that generalizes the determinant-based objective function used by \cite{cacciato2006}.
The specific positions and number of the pixels $\Theta_{\mathrm{crit}}$ are defined by the user in order to keep the method flexible. Depending on the specific dataset under consideration, the \emph{arc pixels} could represent pixels containing giant arcs, the positions of multiple images or model-independent constraints from lensed images \citep[\eg][]{Merten2009,wagner16}. In this study we use pixels randomly sampled over giant arcs, as explained in Section~\ref{demo}. These trace the critical curves to good approximation within the uncertainties of the LTM approach \citep{meneghetti2007}. 
}

\begin{figure}
\centerline{\includegraphics[width=\columnwidth, trim= 0mm 0mm 0mm 0mm,clip]{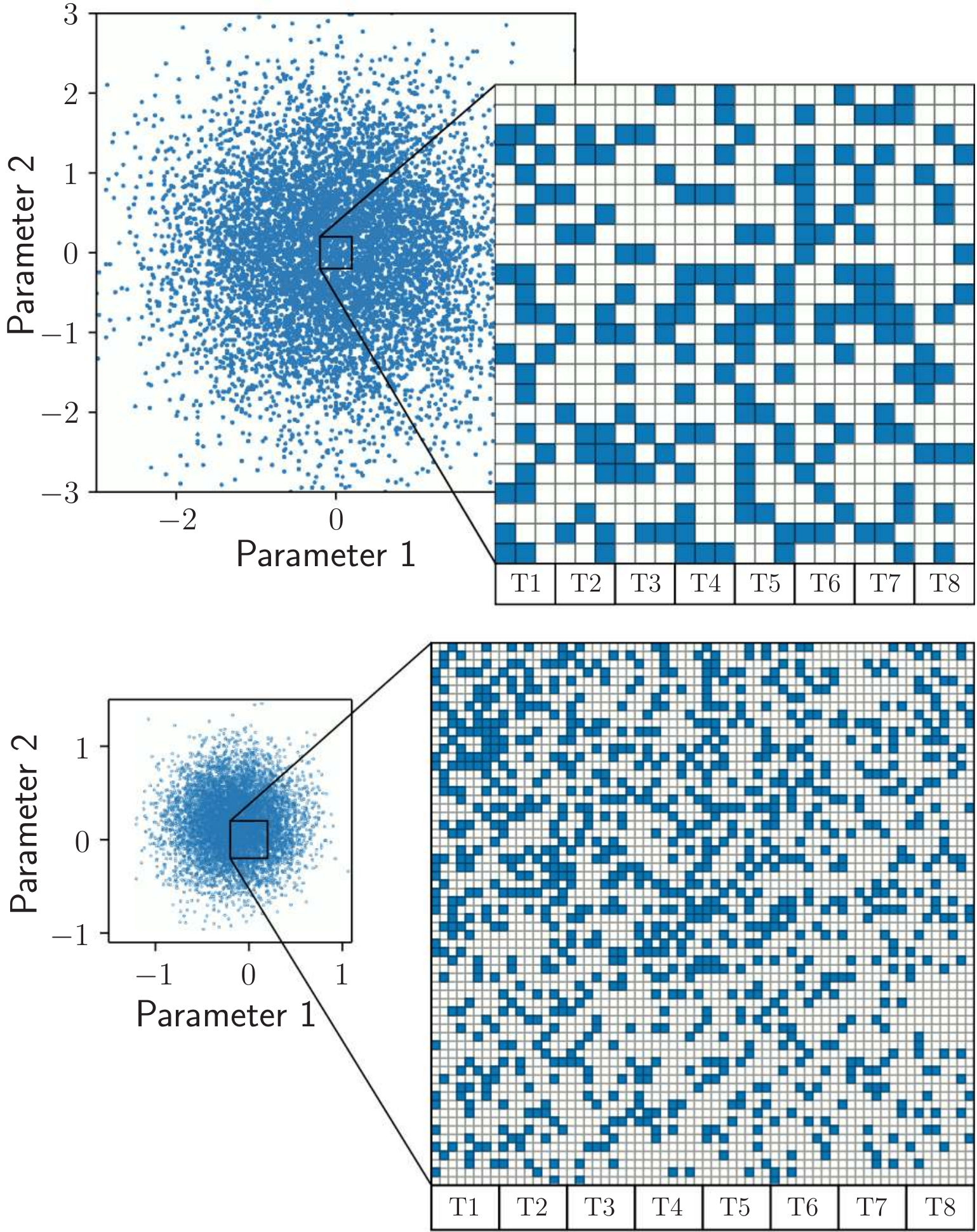}}
\caption{\label{fig:mcmc1}
        Illustration of the adaptive grid-based MCMC method for a two-dimensional parameter space: Gaussian random variates are sampled on a fine regular grid (\emph{top}), which is iteratively refined in size and resolution (\emph{bottom}) according to the previous optimum. The computation is performed columnwise, distributed on a given number of threads, in this example eight (T1 - T8).
}
\end{figure}

For an efficient minimization procedure, we have developed a parallelized generalization of the Metropolis-Hastings algorithm \citep{metropolis, hastings}, which is able to perform up to ten $\smash{\chi^2}$-evaluations per second by combining Markov Chain Monte-Carlo (MCMC) sampling with an innovative use of adaptive grid methods. In particular, the randomly drawn variates for $q$ and $\sigma$ characterizing the shapes of the kernels $G$ and $C$ are binned on a regular grid (\cf Fig.~\ref{fig:mcmc1}) in order to re-use intermediate results between the sampling steps, while retaining the Gaussian distribution of the priors.
The parameter space is explored by multiple threads in parallel, which update the mean values of their prior distributions periodically according to the global $\chi^2$-minimum. Around this minimum, the grid can be iteratively refined to achieve a higher accuracy and precision.
With this scheme, the runtime is reduced by orders of magnitude when compared with a standard MCMC approach (see Subsection~\ref{performance} for runtime measurements).
Currently, the calibration can be performed on individual known lenses to obtain independent sets of parameters, which then need to be combined appropriately.

\hl{
In contrast to the other parameters, the critical number density $n_c$ is defined based on the number of LRGs used.
In the current implementation, we keep $n_c$ constant at an empirically determined value of $n_c \approx 65/(15'\times 15')$ per sheet.}

\hl{In an upcoming implementation, we are going to introduce a multi-lens calibration, in which all parameters are fitted to a larger set of known lenses simultaneously, optimizing the detection rate and the number of predictions to obtain the most efficient set of parameters, which is able to provide the best description for a broad range of lensing mass distributions.
}

\subsection{Performance} \label{performance}

Due to the large area that needs to be explored with the calibrated parameter sets, as well as the large number of MCMC sampling steps to be evaluated during the calibration itself, an optimization of the code-design for cost-effiency, \ie for a low consumption of time and memory, is of critical importance. For the cost-efficiency to scale well with both the number of parallel processes and the problem size, we aim at a maximal parallel portion\footnote{Defined as the proportion of parallelizable workload.} \citep{amdahl1967} and weak scalability.
In our implementation, we apply these criteria by using discrete fast Fourier methods for the convolutions and by parallelizing the computation of lensing quantities using the \texttt{OpenMP}\footnote{\emph{Open Multi-Processing} (\url{http://openmp.org/}).} interface. As an option, specific instructions can be accelerated on a \texttt{CUDA}\footnote{\url{https://developer.nvidia.com/cuda-zone}.}-capable GPU, if available.

In Appendix~\ref{appendix_speedup}, we present runtime measurements for the calibration- and application phases on a recent 64-bit consumer PC\footnote{The test machine runs on a $4.20 \, \mathrm{GHz}$ \texttt{Core i7-7700K} quad-core processor with $32 \, \mathrm{GB}$ RAM ($2400 \, \mathrm{MHz}$ \texttt{DDR4}).} when \emph{EasyCritics} is applied to a realistic scenario and using different numbers of CPU cores. We compare and discuss these results in the context of strong and weak scalability. The typical runtime for a single calibration step remains below a second even in single-thread mode, demonstrating the advantages of the adaptive, grid-based MCMC method. The application to a $\smash{1 \, \mathrm{deg^2}}$ region takes less than a minute on a commercial CPU.
In comparison with crowdsourced visual inspection, these timescales represent a significant improvement. The visual inspection of a $\smash{154 \, \mathrm{deg^2}}$ region by \mbox{37 000} volunteers can require up to eight months \citep{sw1}. For upcoming $\smash{\gtrsim 10^4 \, \mathrm{deg^2}}$ surveys, such an analysis would require the same number of people to inspect the images for half a lifetime, highlighting the high costs and low efficiency of visual inspection. In contrast, \emph{EasyCritics} is able to analyze $\smash{10^4 \, \mathrm{deg^2}}$ on timescales of days or weeks\footnote{Depending on the details of the setup} on a consumer PC, once calibrated.

\section{Application examples} \label{demo}

As a first test to investigate and demonstrate the concepts discussed in this work, we apply \emph{EasyCritics} to the \emph{CFHTLenS} dataset \citep{Heymans12, Erben13, Miller2013}.
\emph{CFHTLenS} is a wide-field photometric survey that covers $\smash{154 \,\mathrm{deg}^2}$ and provides optical images
in the five broadband filters $u^*, g', r', i', z'$. With a depth of $m_{\mathrm{lim}} = 25.5$ in the $i'$-band, sub-arcsecond seeing conditions and accurate photometric redshifts, the \emph{CFHTLenS} imaging data and object catalogs are well suited for gravitational lensing studies \citep{Erben13, Hildebrandt2012}. 

For the use with \emph{EasyCritics}, we filter the object catalogs for early-type galaxies by removing all stellar sources and objects with a spectral type\footnote{Where $T_B = 1$ corresponds to E/S0 galaxies, $T_B = 2$ to SBc barred spirals and $T_B \geq 2$ to spiral and irregular galaxies.} $T_B > 1.7$ according to the classification by \cite{CWW80}.
\hl{
We apply a $k$-correction to the fluxes and magnitudes of the galaxies using the template spectra described by \cite{capak04} and the final transmittance curves of the \emph{MegaCam} filters\footnote{\url{http://www1.cadc-ccda.hia-iha.nrc-cnrc.gc.ca/community/CFHTLS-SG/docs/extra/filters.html}}. However, the $k$-correction does not have a strong impact on the results because it is to a large part degenerate with the free model parameter $K$ and, over small redshift ranges, could be absorbed into the calibration. For the computation of cosmological distances, we use a $\Lambda$CDM model with \emph{Planck} 2015 parameter values \citep{planck}.
}

\hl{
We apply \emph{EasyCritics} to a subset of four individual lenses, selecting well studied objects that cover small and large structures. We start by calibrating the parameters on two known lenses, a small group (“A”) and a rich cluster (“B”).
This yields the two independent parameter sets $P_A$ and $P_B$. Applying both sets one after each other, we then test whether \emph{EasyCritics} is able to predict critical curves around two further known lenses “C” and “D”. As a consistency check, we perform also a reverse test by calibrating the model parameters against the lenses C and D and applying them to the lenses A and B. Moreover, we apply \emph{EasyCritics} to a $\smash{1 \, \mathrm{deg}^2}$ field in which no group- and cluster-scale lens candidates have been reported yet.
The four calibrations are summarized in Subsection~\ref{calib_examples} and the resulting predictions are presented and discussed in Subsection~\ref{appl_examples}.
In Subsection~\ref{error_estimation}, we estimate how the results are affected by uncertainties with regard to the lens- and source redshifts. An overview over the calibration and detection results is presented in Table~\ref{table:sum_res}.
}

Note that a full assessment of the detection efficiency, using the entire survey area and simulations, is beyond the scope of this work and will be presented in our upcoming papers \citep[][in prep.]{2nd_paper, 3rd_paper}.

\subsection{Selection of known lenses} \label{abcd}

\hl{
The lenses A-D are representative for the group- and cluster scale lenses reported in \emph{CFHTLens}. Three of the lenses (A,B,C) are extracted from the \emph{CFHTLS Strong Lensing Legacy Survey} (\emph{SL2S}, \citealt{Cabanac07, more2012}) and \mbox{one (D)} from the newly reported lenses by \emph{SpaceWarps} (\emph{SW}, \citealt{sw1,sw2}) that has so far been missed by automated methods.
The four examples are chosen to “challenge” \emph{EasyCritics} by including lenses with very different properties:
(1) The lenses are located in areas with different depths and environmental properties; (2) The lenses represent different regimes in Einstein radii, masses and redshifts; (3) The lenses have different numbers of arcs, ranging from 1 to 3 arcs/images. Below, we briefly introduce all four objects and describe their main characteristics:
}

\begin{itemize}
	\item \emph{Lens A:} The first object is a compact group of galaxies, \emph{SL2S J021408-053532} or \emph{SA22}\footnote{Referring to the object IDs defined by \cite{more2012} and \cite{sw2}.}, located at $z_{\mathrm{phot}} = 0.48 \pm 0.02 $ \citep{more2012}. 
	It shows a fold configuration of two arcs (A1, \cf Fig.~\ref{fig:lens_closeup}) at $z_{\mathrm{spec}} = 1.628 \pm 0.001$ and a possible third feature (A2, \cf Fig.~\ref{fig:lens_closeup}) observed at $z_{\mathrm{spec}} = 1.017 \pm 0.001$ \citep{verdugo11, verdugo16}. The apparent Einstein radius amounts to $\theta_{\mathrm E} \sim 7.1"$. 
	A combined dynamical and parametric lensing mass reconstruction carried out by \citealt{verdugo16} suggests that the centers and orientations of the main dark matter halo, the intracluster gas halo and the luminosity contours are in good agreement if fitting an NFW profile for the main group halo and pseudo-isothermal ellipses to the central galaxy halos.\\
	
	\item \emph{Lens B:} The second object is the galaxy cluster \emph{SL2S J141447+544703} or \emph{SA100} \citep{Cabanac07} at $z_{\mathrm{phot}} = 0.63 \pm 0.02$ \citep{more2012}, which is known to be a Sunyaev-Zel'dovich bright cluster (\emph{PSZ1 G099.84+58.45}, \citealt{planck13_sz,gruen2014}). It shows a tangential giant arc (B1, \cf Fig.~\ref{fig:lens_closeup}), whose redshift, however, has not yet been measured spectroscopically. With an apparent Einstein radius of $\theta_{\mathrm E} \sim 14.7"$, the cluster belongs to the largest strong lenses imaged by \emph{CFHTLenS}.\\

	\item \emph{Lens C:} The third object is the galaxy group \emph{SL2S J08544-0121} or \emph{SA66} at $z_{\mathrm{spec}} = 0.351$ \citep{Limousin09_SL_Ggroup}, which stands out by a bimodal light distribution with luminosity peaks separated by $\smash{\gtrsim 50''}$. One of these peaks is associated with a bright, double-cored galaxy, near which a typical cusp configuration at \mbox{$z_{\mathrm{spec}} = 1.2680 \pm 0.0003$} \citep{limousin2010_einstein_radius} is observed. The latter shows three images merging into a giant arc and a bright counter image (\cf Fig~\ref{fig:res1}). 
	The lens has an Einstein radius of $4.8"$ and appears to be dominated by a single massive galaxy, which makes this candidate an ideal application target for testing the limits of \emph{EasyCritics} towards smaller scales.\\

        \item \emph{Lens D:} The fourth object is the small cluster \emph{J220257+023432} or \emph{SW7} at redshift\footnote{Estimated by the mean photometric redshift of the central galaxies according to the \emph{CFHTLenS} catalogs.} \mbox{$z_{\mathrm{phot}} \sim 0.51$}, which shows two very promising, although partially faint and overblended giant arc candidates (\cf Fig.~\ref{fig:res1}) at a radius $\theta_{\mathrm{E}} \sim 6.8"$. These have been newly reported as secure detections by \emph{SW} \citep{sw2} after they were missed by various automated algorithms applied within the region \citep[\eg][]{more2012}. The strongly-lensing system is an interesting case to test the potential of the arc-free search approach underlying \emph{EasyCritics}, since a successful detection and identification of this object as a strong lens would be the first achieved with a fully automated method.
       
\end{itemize}

\subsection{Calibrations} \label{calib_examples}

In the following, we describe the four independent calibrations on lenses A-D and discuss their results.
For each lens, the model parameters are estimated via the MCMC-based critical curve fitting method described in Subsection~\ref{calib}. In case that there is more than one configuration of known arcs, we select that covering the largest area.
Where available, we use spectroscopic information on the source redshift.
For the initial parameter values and ranges, we apply empirically motivated values, which are specified in Table~\ref{table:initial_params}. The MCMC sampler is initialized with a Gaussian random distribution that has a width $\sigma_r = 0.4 \Delta_r$, with $\Delta_r$ denoting the interval considered for the respective parameter. 

\begin{figure}
\centerline{\includegraphics[width=\columnwidth, trim= 0mm 0mm 0mm 0mm,clip]{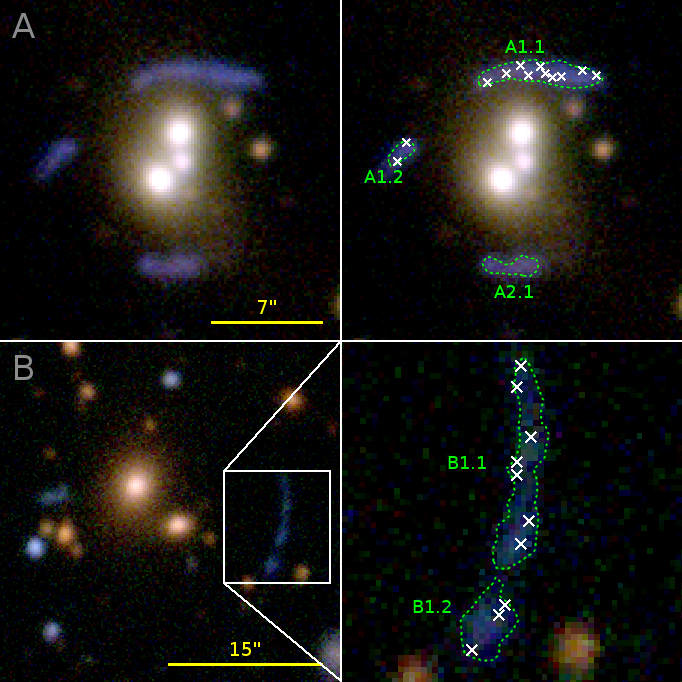}}
\caption{\label{fig:lens_closeup}
        Illustration of the calibration for lenses A and B. The figure shows \emph{CFHTLenS} $i' r' g'$-composite images
        with the arc regions (green contours) and the random $\chi^2$-minimization pixels $\Theta_{\mathrm{crit}}$ (white crosses) used in Subsection~\ref{calib_examples}.
}
\end{figure}

\hl{
The positions of the pixels $\Theta_{\mathrm{crit}}$ to be used for the eigenvalue minimization are sampled randomly over the areas covered by the arcs, as illustrated in Fig.~\ref{fig:lens_closeup}. These areas are defined by a visual selection of suitable $g'$-band contours. As discussed in Subsection~\ref{calib}, the true path of the critical curve is in principle unknown, but can be approximated to good accuracy by the positions of the arcs for an estimation of the LTM model parameters.
The uncertainty $\Delta \chi^2$ is estimated based on the largest difference in the Jacobian eigenvalues when randomly varying the location of the reference pixels within the selection area.
The final outcome of the calibration is given by the set sitting in the global minimum within $\Delta \chi^2$.
}

We present the four results $P_A$ to $P_D$ together with their $1\sigma$-uncertainties due to $\Delta \chi^2$ in Table~\ref{table:res1_params} and show the associated best-fitting critical curves in Fig.~\ref{fig:res1} and Fig.~\ref{fig:res2}.
The fits converged within $\smash{N \lesssim \mathcal O(10^5)}$ sampling steps, resulting in final least-square deviations between $\smash{\chi^2 \sim \mathcal O(10^{-3})}$ and $\smash{\chi^2 \sim \mathcal O(10^{-2}})$.
The calibrated critical curves qualitatively match our expectation. For instance, the result for lens A resembles the lens modelling reconstruction of \mbox{\cite{verdugo11}} well. In Fig.~\ref{fig:res1}, we compare the critical curve obtained with our approach (solid line) and the result of Verdugo et al. (dashed line).
A slight deviation can be noted in the bottom image region, but this is well explained by a slight misalignment between the smoothed $r'$-band luminosity isocontours and the total mass. 
For lens D, the shape of the predicted critical curve might have been affected slightly too much by local galaxies in the proximity of the northern giant arc candidate. However, its location and size match the arcs well.

\begin{table}
	\centering
	\caption{Initial parameter values and ranges considered for the calibration, where $K$ is given in arbitrary units.}
        \label{table:initial_params}
	\setlength\tabcolsep{12pt}
	\begin{tabular*}{\columnwidth}{rcccc}
		\toprule
		\toprule
		& $q$ & $K$ & $\mu_{\mathrm{clus}}$ & $\sigma$ [$''$]\\
                \midrule
                Lower limit: & $1.10$ & $\hphantom{10}5$ & $0.70$ & $\hphantom{1}3$ \\
                Initial value: & $1.25$ & $\hphantom{1}47$ & $0.85$ & $10$\\
                Upper limit: & $1.40$ & $100$ & $0.95$ & $20$\\
		\bottomrule
	\end{tabular*}
\end{table}

\hl{
A quantitative comparison of the best-fitting parameter values for the lenses $P_A$ to $P_D$ listed in Table~\ref{table:res1_params} shows that the scatter in $q$ and $K$ between the lenses within the estimated uncertainties is low despite the large difference in the redshift and Einstein radii of the lenses. 
As expected, the largest differences between all four parameter sets are found between lenses $B$ and $C$.
Despite a possible impact of the low number of satellites on the $M/L$ estimate in the case of lens C \citep{lu15}, the different values obtained for $K$ are consistent with the expected scatter in the $M/L$ ratios of groups and clusters \citep[\eg][]{lin2004, tinker2005, bahcall2014, lu15, viola15, mulroy17}. 
}

\begin{table}
        \centering
        \caption{Calibration results for all four lenses A-D. The parameter $K$ is given in arbitrary units.}
        \label{table:res1_params}
        \setlength\tabcolsep{8pt}
        \begin{tabular*}{\columnwidth}{rcccc}
                \toprule
                \toprule
                Lens & $q$ & $K$ & $\mu_{\mathrm{clus}}$ & $\sigma$ [$''$]\\
                \midrule
                A & $1.24 \pm 0.02$ & $37\pm7$ & $0.84\pm0.01$ & $10 \pm 1$ \\
                B & $1.22 \pm 0.02$ & $35\pm7$ & $0.93\pm0.01$ & $18\pm1$ \\
                C & $1.28 \pm 0.01$ & $68\pm6$ & $0.81\pm0.01$ & $\hphantom{1}5\pm1$ \\
                D & $1.26 \pm 0.02$ & $40\pm7$ & $0.81\pm0.02$ & $10\pm1$ \\
                \bottomrule
        \end{tabular*}
\end{table}

\subsection{Identification of known lenses} \label{appl_examples}

\hl{
For each of the four lenses, we now apply \emph{EasyCritics} to a $5' \times 5'$ (extended: $15' \times 15'$) region around the cD galaxies. The precise center of the region is not relevant because in a general application, the algorithm is run over the entire dataset. For a positive detection, we require that at least one of the parameter sets applied produces a critical curve with a centroid separated by less than $40''$ from the expected lens center\footnote{Defined as the predicted maximum $\boldsymbol{\theta}_{\mathrm{c}}$ of the convergence within the region enclosed by the visible arcs.}. The choice of this maximum distance accounts for intrinsic offsets between the dark and luminous matter components. 
In addition, we investigate the agreement of the predicted critical curves with those obtained from the calibrations, which represent the best fit of our LTM model. This allows to study the impact of the intrinsic variation of lens parameters on the model predictions.
}

\hl{
In general, the variation of parameters from lens to lens is expected to yield slight deviations of the predicted Einstein radii from the true ones. These, however, are well within the aims of lens detection. In fact, the goal is not a precise reconstruction of the mass, which can be achieved only via a subsequent fit, but to find the regions in which strong lensing features are produced.
Thus, the sensitivity of the predicted critical curves to variations in the model parameters should become relevant only to the extent to which it may affect the detection rate or the area for follow-up inspection.
An extensive analysis of the detection efficiency of \emph{EasyCritics} for different settings will be the subject of upcoming papers \citep[][in prep.]{2nd_paper}.
Here, we focus on the individual performance of the parameter sets $P_{A}$ to $P_{D}$.
}

The tests yield positive outcomes for all four lenses. Each lens is identified by each parameter set, except for lens C, for which a critical curve is produced only by the group-scale set $P_A$. This, however, is not surprising given the very low richness of lens C, for which $P_A$ may provide a better description than $P_B$.
A more remarkable result is the successful identification of the \emph{SW} lens candidate D, as it had been missed by several automated methods before \citep{more2012}. This demonstrates the importance and benefits of using foreground search criteria complementary to arc data, thus avoid problems due to low signal-to-noise ratios, complex morphologies or overblending.

\begin{figure}
\centerline{\includegraphics[width=\columnwidth, trim= 0mm 0mm 0mm 0mm,clip]{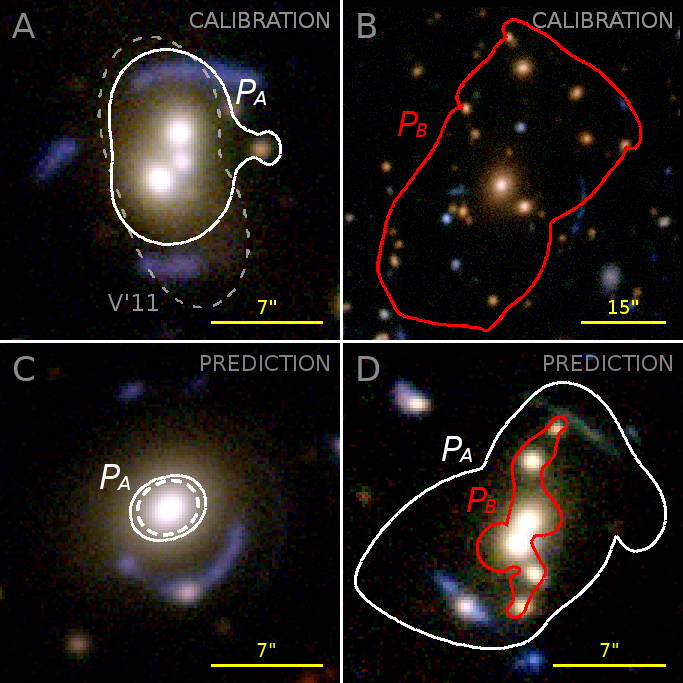}}
\captionof{figure}{\label{fig:res1}
        Calibrated and predicted tangential critical curves for the first test scenario, shown on \emph{CFHTLenS} $i'r'g'$-composite images. \emph{Top row:} Calibrations $P_A$ and $P_B$ on the two lenses A and B (solid lines) and literature comparison showing the model of \citet[][V'11]{verdugo11}. \emph{Bottom row:} Predictions for lenses C and D using $P_A$ and $P_B$ and assuming $z_s = 2$ (solid lines) and for lens C the known value $z_s = 1.268$ (dashed line).\vspace{-7pt}}
\end{figure}

\begin{figure}
\centerline{\includegraphics[width=\columnwidth, trim= 0mm 0mm 0mm 0mm,clip]{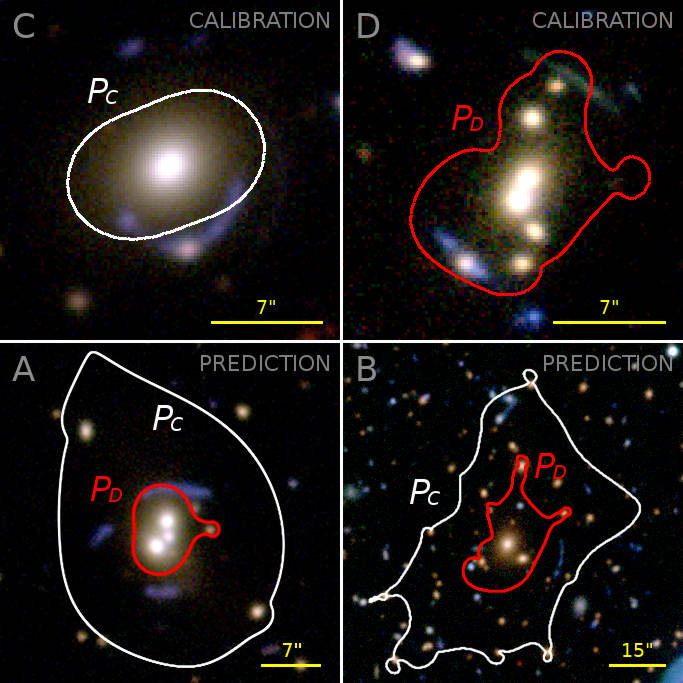} \hfill}
\caption{\label{fig:res2}Calibrated and predicted tangential critical curves for the second test scenario. \emph{Top row:} Calibrations $P_C$ and $P_D$ on the lenses C and D. \emph{Bottom row:} Tangential critical curves predicted for lenses A and C by $P_C$ and $P_D$, assuming $z_s = 2$ where the source redshift is not known.}
\end{figure}

\begin{figure}
\centerline{\includegraphics[width=\columnwidth, trim= 0mm 0mm 0mm 0mm,clip]{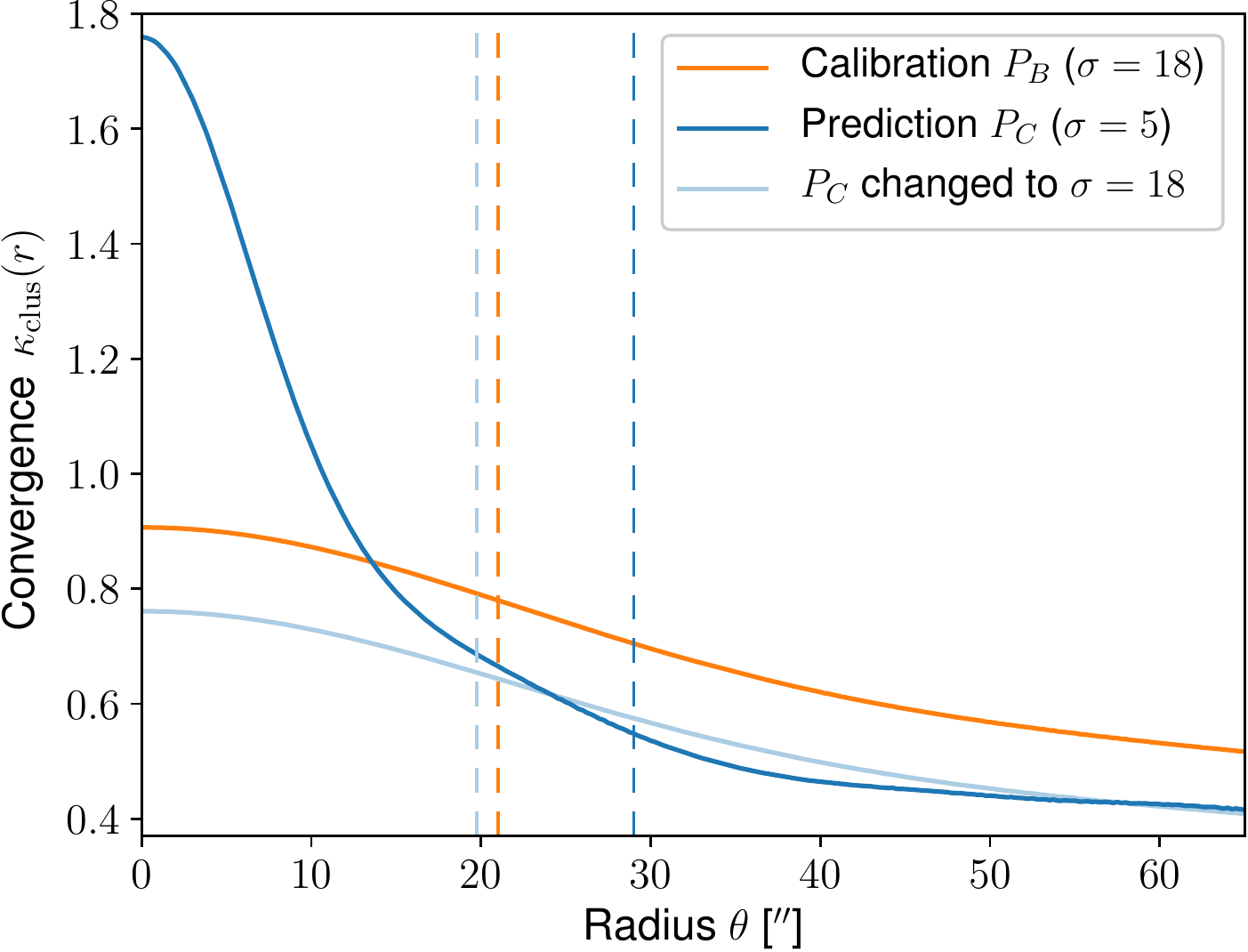} \hfill}
\caption{\label{fig:damping}
Comparison of the convergence profile $\bar\kappa_{\mathrm{clus}}(r)$ for the smooth cluster-scale component of lens B, illustrating the behavior of different sizes $\sigma$ for the Gaussian smoothing kernel.
The curves show the profiles computed for $P_B$, $P_C$ and when using $P_C$ with the value $\sigma$ from $P_B$. The dashed lines mark the predicted effective Einstein radii $\theta_{\mathrm{E,eff}}$. \vspace{-5pt}}
\end{figure}

\hl{
A comparison of all four predicted critical curves is shown in Fig.~\ref{fig:res1} and Fig.~\ref{fig:res2}. Not surprisingly, the parameters obtained from small/large lenses perform best on the other small/large lenses.
At the same time, the smallest-scale calibrations $P_C$ and $P_D$ show an overall better performance than the cluster-scale calibration $P_B$ by predicting critical curves with either large or compatible Einstein radii for all mass regimes, while $P_B$ strongly underestimates the surface density on group scales.
This is because a small Gaussian width $\sigma$ (resulting from the calibration on a small lens) applied to large lenses is still going to result in cluster-scale halos concentrated enough to produce critical curves close to the Einstein radii, which is no longer the case when applying a large $\sigma$ to small lenses. In this case, the predicted cluster-scale component is spread over a large area producing an excessive dilution that can prevent the appearance of critical regions.
This effect is illustrated in Fig.~\ref{fig:damping} for the example of lens B, where it gets further enhanced by the fact that multiple halos are superposed. In contrast, the remaining parameters $q,K,\mu_{\mathrm{clus}}$ have a much smaller influence on the model. This would allow to fix fiducial values for these parameters and to perform an optimization on the parameter $\sigma$ in order to reduce the degrees of freedom of the model, at least for the sample of lenses studied here.
}

\begin{table*}
        \centering
        \caption{Summary of quantitative results from Subsections~\ref{calib_examples} and \ref{appl_examples}.}
        \label{table:sum_res}
        \begin{tabular*}{\textwidth}{lcccllcl}
                \toprule \toprule
                {Lens}
        &{\begin{tabular}{@{}c} Right ascension\\ $\alpha$ (J2000) $[\mathrm{deg}]$ \end{tabular}}
        &{\begin{tabular}{@{}c} Declination $\delta$\\ (J2000) $[\mathrm{deg}]$  \end{tabular}}
        &{$\begin{array}{@{}l} \text{Lens red-}\\\text{shift } z_l \end{array}$}
        &{\begin{tabular}{@{}c} Detected?\\(No. of\\curves) \end{tabular}}
        &{$\begin{array}{@{}l} \text{Arc ra-}\\\text{dius } \theta_{\mathrm{E}} \text{ ['']} \end{array}$}
        &{$\begin{array}{@{}l} \text{Calibrated}\\ \text{eff. Einstein}\\ \text{radius } \theta_{\mathrm{E,eff}} \text{ ['']} \end{array}$}
        &{$\begin{array}{@{}l} \text{Predicted}\\ \text{eff. Einstein}\\ \text{radii } \theta_{\mathrm{E,eff}} \text{ ['']} \end{array}$}\\
                \midrule
                A & 33.5336 & $-$5.5922 & 0.48\hphantom{0} & Yes (2) & \hphantom{0}7.1 & 5 & $P_C$: 15; \; $P_D$: 4 \\
                B & 213.6965\hphantom{0} & $+$54.7842\hphantom{0} & 0.63\hphantom{0} & Yes (2) & 14.7 & 21 & $P_C$: 29; \; $P_D$: 12 \\
                C & 133.6940\hphantom{0} & $-$1.3603 & 0.351 & Yes (1) & \hphantom{0}4.8 & 5 & $P_A$: 2 \\
                D & 330.7372\hphantom{0} & $+$2.5760 & 0.51\hphantom{0} & Yes (2) & \hphantom{0}6.8 & 6 & $P_A$: 8;\hphantom{0} \; $P_B$: 3 \\
                $1 \, \mathrm{deg}^2$ field & 215.3273\hphantom{0} & $+$55.2480\hphantom{0} & 0.3\dots0.4 & - & \hphantom{0.} - & - & $P_C$: 3\\

                \bottomrule
        \end{tabular*}
\end{table*}

%-------------------
\begin{figure*}
\begin{center}
\includegraphics[width=\textwidth]{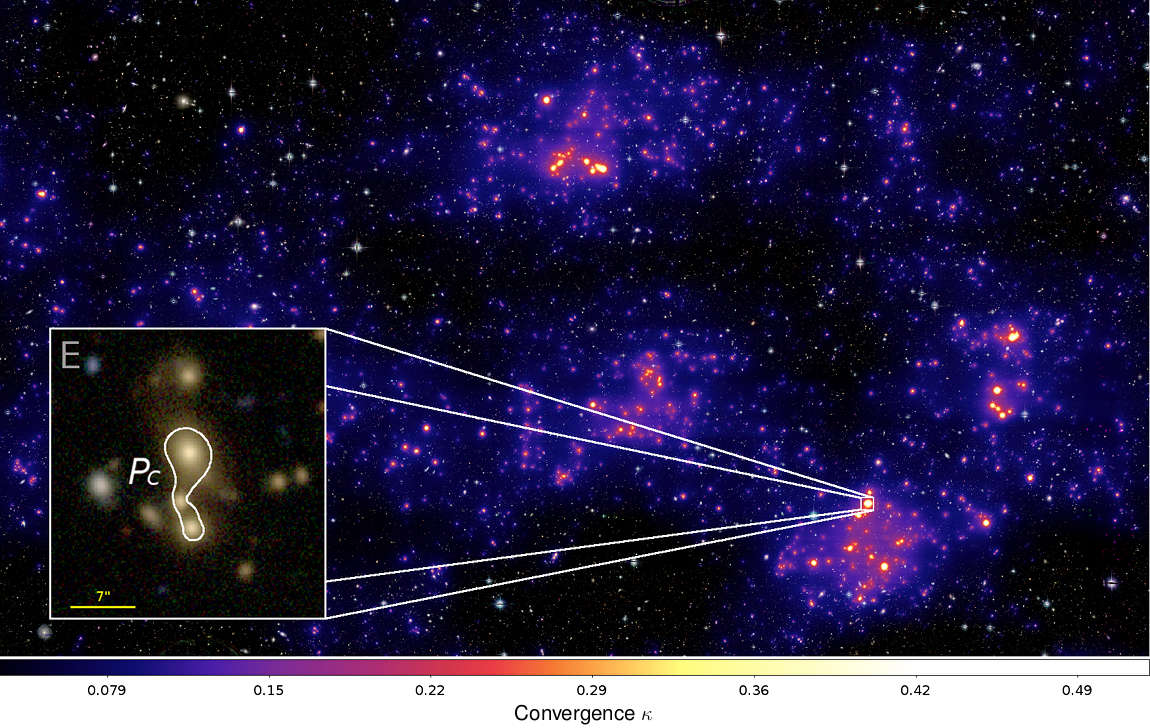} \vspace{-10pt}
\caption{\label{fig:vt1}
        Cluster-scale convergence predicted with parameter set $P_C$ for the candidate-free $1 \, \mathrm{deg}^2$ region, superimposed on a \emph{CFHTLenS} $i' r' g'$-composite image.
        The box on the left shows a zoom-in on the critical curve of the only group-scale critical curve detection.
}
\end{center}
\end{figure*}
%-----------------

\hl{
In the remaining part of our tests, we apply \emph{EasyCritics} to a $\smash{1 \, \mathrm{deg}^2}$ field
for which no group- or cluster-scale strong lens candidates have been reported by the \emph{SL2S} \citep{Cabanac07,more2012}, by \emph{SW}\footnote{As of November 10, 2015.} \citep{sw2} or in \emph{NED}\footnote{\emph{NASA/IPAC Extragalactic Database}, \url{https://ned.ipac.caltech.edu/}.}. In contrast to the previous examples, we display a case for a low density region in which we expect to find either new or no critical regions. 
In this field, \emph{EasyCritics} produces only a single critical curve on group- or cluster scales with $\theta_{\mathrm {E,eff}} \approx 3"$, found by the set $P_C$ and centered around what appears to be an elongated small cluster (“E”). The latter can be estimated to be at $z \sim 0.3\dots0.4$ based on the photometric redshifts of the elliptical galaxies in the region.
In addition, very few critical curves are produced around single isolated field galaxies, all having Einstein radii $\theta_{E,\mathrm{eff}} < 2.75"$. These features are mostly artificial detections due to the central cusp in our description of galaxies, which can give rise to few pixels with supercritical surface densities.
They can be avoided easily by increasing the critical curve size threshold or by a slight modification of the parameter values. For instance, in configuration $P_B$, only one of these objects is created.
A map of the convergence for the entire region and a zoom-in on the central region of candidate E is presented in Fig.~\ref{fig:vt1}.}

\subsection{Impact of redshift uncertainties} \label{error_estimation}

Apart from the intrinsic variation in the best-fitting parameters, some impacts on the models may arise due to the photometric redshift measurement uncertainty, $\Delta z_{\mathrm{phot}}$, and the intrinsic scatter of the source redshifts, $\Delta z_s$. Both could mildly affect the detection statistics by altering the calibration and the predicted sizes of critical curves.
In order to estimate at least the order of magnitude of these effects, we study how the critical curves are affected by small displacements in the lens- and source redshifts and compare this to the previously discussed sources of uncertainty. The results obtained in this way are very similar for all four lenses and in the following discussed in detail for the example of the cluster-scale lens B, which is farthest away in redshift and thus may be affected the most by the photometric uncertainties.

\begin{figure}
\centerline{\hfill\includegraphics[width=\columnwidth, trim= 0mm 0mm 0mm 0mm,clip]{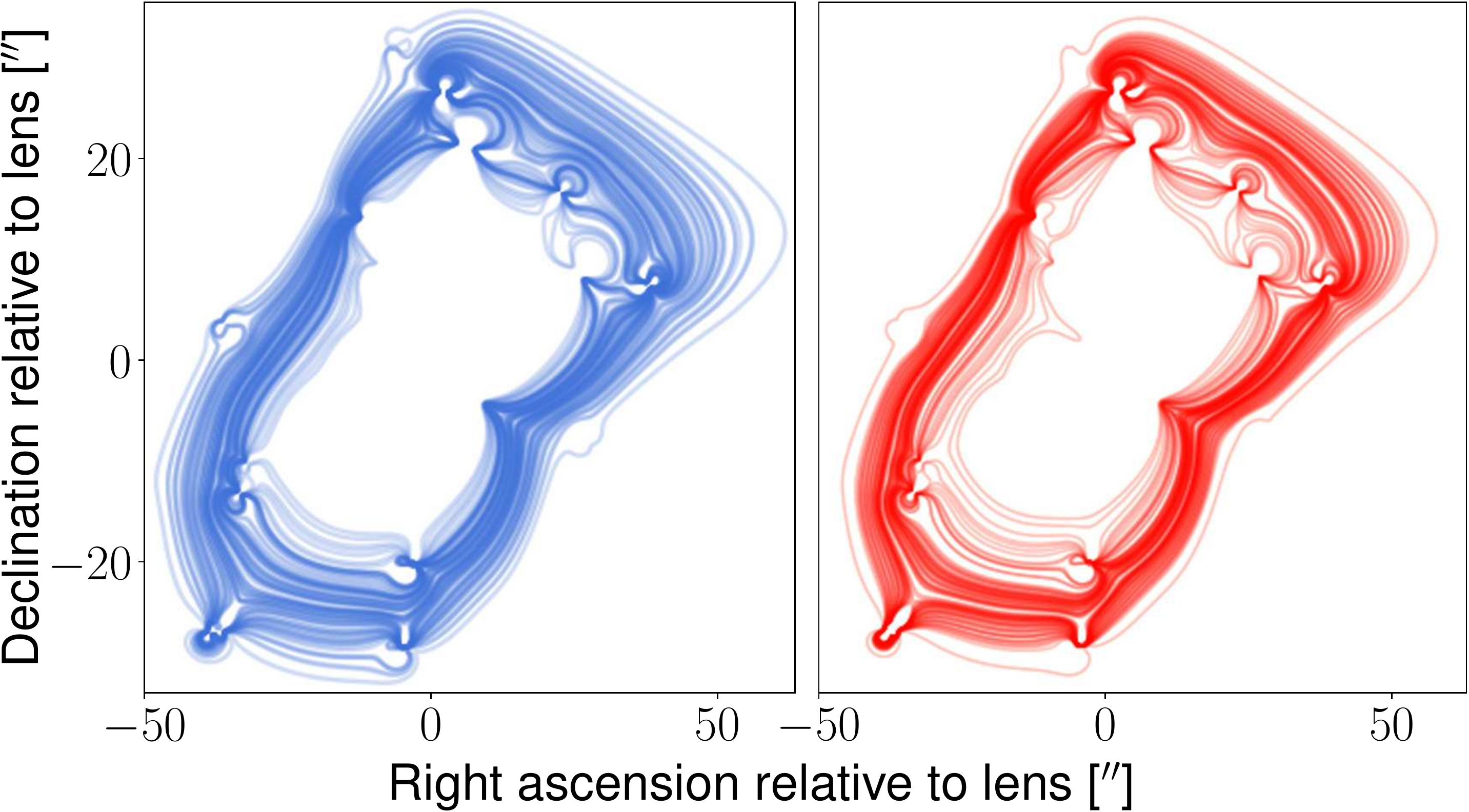}}
\caption{\label{fig:z_error}Impact of the mean photo-$z$ uncertainties (\emph{left/blue}) and the source redshift uncertainties (\emph{right/red}). Shown are in both cases the tangential critical curves of lens B produced with $P_B$ for different random displacements of the redshifts, as described in Subsection~\ref{error_estimation}. For a clearer presentation, we only show every second of the 200 lines}
\end{figure}

We first investigate the sensitivity of predicted critical curve sizes to the average photometric measurement uncertainty $\Delta z_{\mathrm{phot}} = 0.04(1+z)$ of the \emph{CFHTLenS} data (\cf \citealt{Hildebrandt2012}). We displace all redshift bins by Gaussian random variates with a mean $\mu_r = z^{(k)}$ and a standard deviation $\sigma_r = \Delta z_{\mathrm{phot}}$, while keeping the parameters fixed at the calibration set $P_B$. The procedure is performed 200 times, yielding the critical curves shown in the left panel of Fig.~\ref{fig:z_error}. The mean Einstein radius is $\theta_{\mathrm{E,eff}} = 21"$ with a standard deviation of $2.6"$.
The results confirm our expectation that the photometric measurement uncertainties influence the sizes of the critical curves mildly, although leaving the overall outcome unchanged. 
We repeat the calibration for the random displacements that yielded critical curve sizes closest to the $1\sigma$ neighborhood around the mean to estimate the impact of $\Delta z_{\mathrm{phot}}$ on the parameter values. The results are shown in Table~\ref{table:res1_errorbars}.

In a second test, we investigate the change in the calibrated and predicted critical curves when the source redshift differs from our default expectation value $z_s = 2$. This does not represent a test of the lens model itself, but of its dependency on the source redshift only.
We first study the change in predicted critical curves by re-applying the parameter values $P_B$ calibrated under the assumption $z_s = 2$ to \mbox{lens B}, using 200 random redshift displacements within the empirically motivated range of bright giant arcs, $z_s = 2.0 \pm 0.1$ \citep{Bayliss12}. 
The critical curves predicted for these different source redshifts are visualized in the right panel of Fig.~\ref{fig:z_error}.
The effective Einstein radii are distributed with a mean of $\theta_{\mathrm{E,eff}} = 21"$ and a standard deviation of $2"$.
As before, we study also the impact on the calibration by repeating the calibration for the various source redshifts, which yields the maximum differences $\Delta z_s$ in the final best-fitting parameters $P_B$ listed in Table~\ref{table:res1_errorbars}.

\begin{table}
        \centering
        \caption{Comparison of absolute uncertainties in the calibrated parameter values due to (1) redshift misestimations, (2) the degeneracy between optimization points and (3) the intrinsic deviation of $P_B$ from $P_A$.}
        \label{table:res1_errorbars}

        \begin{tabular*}{\columnwidth}{rcccc}
                \toprule \toprule
                {\begin{tabular}{@{}r}Para-\\meter\end{tabular}}
                &{\begin{tabular}{@{}c}Dev. due \\to $\Delta z_{\mathrm{phot}}$ \end{tabular}}
                &{\begin{tabular}{@{}c}Dev. due\\to $\Delta z_s$ \end{tabular}}
                &{\begin{tabular}{@{}c}Dev. due\\to $\Delta \chi^2$\end{tabular}}
                &{\begin{tabular}{@{}c}Dev.\\from $P_A$\end{tabular}}\\
                \midrule
                $q$ & 0.05 & 0.01 & 0.02 & 0.02\\
                $K$ & \hphantom{0.}16 & \hphantom{0}4.5 & \hphantom{0.0}7 & \hphantom{0.0}2\\
                $\mu_{\mathrm{clus}}$ & 0.04 & 0.01 & 0.01 & 0.09 \\
                $\sigma$ & \hphantom{0.0}2 & \hphantom{0.0}2 & \hphantom{0.0}1 & \hphantom{0.0}8\\
                \bottomrule
        \end{tabular*}
\end{table}

The tests show that the detection of the critical curve itself is stable with respect to the redshift uncertainties $\Delta z_{\mathrm{phot}}$ and $\Delta z_s$, although the critical curve may vary in size. In extreme cases, this might affect the completeness of predictions due to underestimating the size of critical curves.

\section{Summary and conclusions}

We have developed an algorithm to detect group- and cluster-scale strong lenses in wide-field surveys, based on the distribution of fluxes from LRGs as a tracer for dense structures.
\emph{EasyCritics} models these structures in a blind and automated way to identify the most likely strongly-lensing groups and clusters of galaxies. 
This promises to avoid a large number of spurious detections and to decrease the required amount of post-processing, as it significantly reduces and restricts the area that needs to be validated or inspected by a follow-up study.
The main advantage of \emph{EasyCritics} is that it enables a simple, physically motivated selection criterion based on few model assumptions and without relying on the identification of arcs.
In addition, \emph{EasyCritics} provides a first characterization of the properties of lens candidates, such as their positions, surface densities, Einstein radii and orientations of the critical curves.

In this work, we presented a modified and numerically improved version of the LTM modelling approach, based on the brightest early-type galaxies distributed along the line of sight. In particular, \emph{EasyCritics} blindly models the lensing distortion produced within the entire line of sight (not only pre-selected, isolated structures) and derives lensing quantities from a potential instead of working with the deflection angle. In combination with the use of parallelization and fast Fourier and matrix calculation routines, this allows to obtain a substantial speed-up in the identification and modelling of strongly-lensing structures.
The optimization of time and memory resources is important to enable the analysis of very large portions of the sky and to automate the estimation of the model parameters based on the adaptive MCMC methods we introduced in this work.

We applied \emph{EasyCritics} to regions centered on known lens candidates to provide a detailed description of actual cases. In an analysis of the primary uncertainties, we showed that the accuracy of the predicted critical curves returns stable results.
In one of the presented examples, the lens candidate detected by \emph{EasyCritics} had previously been missed by automated searches and was eventually discovered through crowdsourced visual inspection. Its successful identification by \emph{EasyCritics} is very encouraging, considering that it required no community efforts and that the overall area of \emph{CFHTLenS} can be processed with two parameter sets in a few hours to days.
The result demonstrates the power of combining lens detection with LTM modelling for an improved automated search based on foreground observables.
A full assessment of the purity and completeness of the catalog of strong lensing features is going to be presented in the two forthcoming papers, in which \emph{EasyCritics} will be applied to the entire \emph{CFHTLenS} survey and tested against simulations. 
For the near future, we furthermore prepare a release of the \emph{EasyCritics} code, which is going to be made available on request at a suitable time.

\emph{EasyCritics} enables a reliable, fast and automated detection and characterization of strongly-lensing galaxy groups and galaxy clusters in wide-field optical surveys. It will thus be able to contribute to various near-future studies exploring the dark universe.

\section*{Acknowledgements}

We gratefully acknowledge valuable discussions and comments from Matthias Bartelmann, Jenny Wagner, Thomas Erben, Gregor Seidel and Lars Henkelmann.
This work was supported by the \emph{Transregional Collaborative Research Center TRR33 'The Dark Universe'} (MC, MM).

%%%%%%%%%%%%%%%%%%%%%%%%%%%%%%%%%%%%%%%%%%%%%%%%%%



\bibliographystyle{mnras}
%\bibliography{new_refs} % if your bibtex file is called example.bib
%\bibliography{new_refs}
\input{paper.bbl}

%%%%%%%%%%%%%%%%%%%%%%%%%%%%%%%%%%%%%%%%%%%%%%%%%%

%%%%%%%%%%%%%%%%% APPENDICES %%%%%%%%%%%%%%%%%%%%%

%\newpage
\appendix

\section{Redshift dependence of the critical surface density}

\vspace{5pt}
\begin{center}
	\captionsetup{format=A1}
	\includegraphics[width=\columnwidth, trim= 0mm 0mm 0mm 0mm,clip]{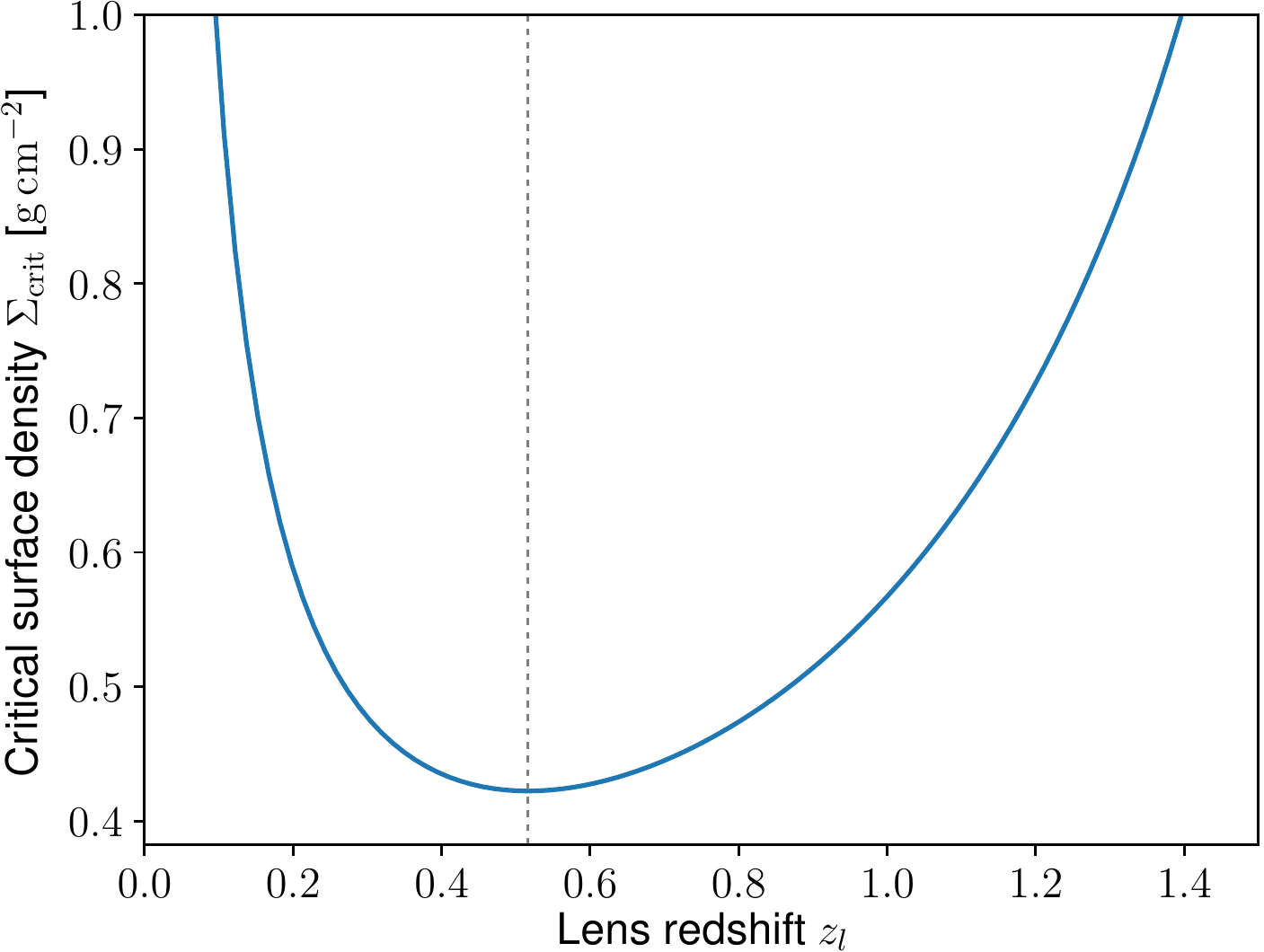}\vspace{-10pt}
    \captionof{figure}{}
    \label{fig:crit_density}
\end{center}

\section{Smoothed cluster kernel} \label{App:AppendixA}

In the following, we briefly outline the steps for solving the convolution integral in Eq.~\ref{final_C},
\begin{align}
        C(\boldsymbol \theta) &= \frac 1{2\pi \sigma^2} \int_{\mathbb R^2} \exp \left(- \frac{\| \boldsymbol \theta - \boldsymbol \theta^{\prime} \|^2}{2 \sigma^2} \right) \| \boldsymbol \theta' \|^{2-q} \, \mathrm d^2 \theta'. \label{G_clus2}
\end{align}
Applying the convenient choice of polar coordinates $(\theta', \phi')$ such that the orientation of the basis vector $\boldsymbol{\mathrm u_{\theta}}$ coincides with the $\phi' = 0$ axis, the angular integration simply yields a Bessel function of the first kind at order zero:
\begin{align}
        C(\boldsymbol \theta) &= \frac {\exp \left( {- \frac{ \theta^2}{2 \sigma^2}} \right)}{\sigma^2} \int_0^{\infty} I_0 \left( \frac{\theta \theta'}{\sigma^2} \right) \exp \left(- \frac{\theta^{\prime 2}}{2 \sigma^2} \right) \theta^{\prime \, 3-q} \; \mathrm d\theta' \label{G_clus3}.
\end{align}

It then turns out convienent to rewrite the Bessel function $I_0$ in terms of the confluent hypergeometric limit function ${}_0F_1$ by exploiting the relation \cite[9.63 \& 9.6.47]{abramowitz64}:
\begin{align}
        I_{0}(z) &= {}_0F_1 \left(; 1; - \frac{z^2}4 \right). \label{bessel1}
\end{align}
This simplifies the integral to:
\begin{align}
        C(\boldsymbol \theta) &= \frac {\mathrm e^{- \frac{ \theta^2}{2 \sigma^2}} }{\sigma^2} \int_0^{\infty} \exp \left({- \frac{\theta^{\prime 2} }{2 \sigma^2} } \right) \theta^{\prime 3-q} \, {}_0F_1\left(;1; \frac{(\theta \theta')^2}{4 \sigma^4} \right)  \mathrm d\theta^{\prime}. \label{intermediate_integral}
\end{align}
Applying the substitutions $\smash{\alpha \equiv 2 - \frac q2}$, $\smash{\beta \equiv 1}$ and $\smash{z \equiv \theta^2/(2 \sigma^2)}$, the remaining expression can be identified as an integral representation of the confluent hypergeometric function $_1F_1$ \cite[6.5.1]{soni1966formulas}:
\begin{align}
        _1F_1(\alpha; \beta; z) = \frac 1{\Gamma(\alpha)} \int_0^{\infty} \mathrm e^{-x} \, x^{\alpha -1} \, {}_0F_1(;\beta; zx) \;\mathrm dx. \;\;
\end{align}
This representation is valid for $\Re(\alpha) > 0$ and thus $q < 4$, which is true for all cases we consider (\cf Subsection~\ref{ml_conv1}).

Furthermore, both the integral and the exponential prefactor in Eq.~\ref{intermediate_integral} can be replaced by a single confluent hypergeometric function if applying Kummer's transformation rule \cite[6.12]{soni1966formulas}:
\begin{align}
        \mathrm e^{x} \, {}_1F_1(\alpha ; \beta ;-x) = {}_1F_1(\beta - \alpha ; \beta ;x).
\end{align}
With this, the result in Eq.~\ref{final_C} follows immediately:
\begin{align}
        C(\boldsymbol \theta) &= \frac{2^{1- \frac q2}}{\sigma^{q-2}} \, \Gamma \left(2- \frac q2 \right) {}_1 F_1 \bigg(\frac q2 - 1; 1; -\frac{\theta^2}{2 \sigma^2} \bigg).
\end{align}

\section{Critical curve shape parameters} \label{appendix_moments}

The orientation $\phi$ and the principal axis lengths $a$ and $b$ are defined by \citep{moments2, stobie86}:
\begin{align}
        a^2 &= 2 \left( \mu_{20} + \mu_{02} \right) + 2 \sqrt{ 4 \mu_{11}^2 + (\mu_{20} - \mu_{02} )^2 }; \\
        b^2 &= 2 \left( \mu_{20} + \mu_{02} \right) - 2 \sqrt{ 4 \mu_{11}^2 + (\mu_{20} - \mu_{02} )^2 }; \\
        \phi &= \frac 12 \arctan \left( \frac{2 \mu_{11}}{\mu_{20} - \mu_{02}} \right).
\end{align}
Here the normalized centered second-order image moment is defined by:
\begin{align}
        \mu_{pq} &= \frac 1A \int_{\partial A} (\theta_1 - \bar \theta_1)^p \, (\theta_2 - \bar \theta_2)^q \; \mathrm d s,
\end{align}
and the centroid is given by:
\begin{align}
        \boldsymbol{\bar \theta} &= \frac 1A \displaystyle \int_{\partial A} \boldsymbol \theta \; \mathrm d s.
\end{align}

\section{Performance and scalability} \label{appendix_speedup}

For the calibration mode, we measured the average durations of single iterations when using the CPU only and when using GPU acceleration. The average is performed over all sampling iterations ($= 10^6$) by measuring the total calibration runtime in seconds and dividing by the number of iteration steps. Note that the speed-up of the iterations due to a usage of multiple CPU cores\footnote{Here, we refer to physical, not logical, cores.} depends on a variety of parameters related to the MCMC sampling procedure. Thus, the values in Table~\ref{table:perf} are presented only to give an impression of the order of the involved timescales; and error bars have been omitted.
The duration of a single iteration needs to be kept as short as possible because a large number of MCMC steps is necessary for a reliable estimation of the model parameters. Most of the acceleration is achieved by the re-use of resources in the adaptive MCMC method (\cf \autoref{calib}), which enables a speed-up by a factor of ten compared to conventional MCMC methods. The remaining gain from the GPU acceleration contributes only a small part of the speed-up for our setup due to the unavoidable bottleneck arising from the data transfer between host and GPU device..

\begin{table}
        \centering
        \caption{Average duration (in miliseconds) of a single MCMC sampling step for a region of $\smash{15' \times 15'}$ as a function of used CPU cores on our test system.}
        \label{table:perf}
        \begin{tabular*}{\columnwidth}{r ccccc} % four columns, alignment for each
                \toprule \toprule
                {Number of CPU cores:}
        &{1}
        &{2}
        &{3}
        &{4}
        &\\
                \midrule
                Time in CPU-only mode $[\mathrm{ms}]$: & $360 $ & $240 $ & $170 $ & $130 $ & \\
                Time in GPU-accelerated mode $[\mathrm{ms}]$: & $280 $ & $190 $ & $140 $ & $110 $ &\\
                \bottomrule
        \end{tabular*}
\end{table}

\begin{table*}
        \centering
        \caption{Average runtimes (in seconds) and their statistical uncertainties for application runs on our test system for different region sizes as a function of used CPU cores.}
        \label{table:perf2}

        \begin{tabular*}{\textwidth}{r D{,}{\,\pm\,}{-1} D{,}{\,\pm\,}{-1} D{,}{\,\pm\,}{4.4} D{,}{\,\pm\,}{4.4} }
        \toprule \toprule
        {No. of CPU cores:}
        &{1}
        &{2}
        &{3}
        &{4}\\
        \midrule
        Time for $\smash{15' \times 15'}$ (4 tiles, not extended) $[\mathrm{s}]$:  & 5.81,0.18 & 3.69,0.13 & 2.84,0.01 & 2.596,0.003\\ %
        Time for $\smash{\,1 \, \mathrm{deg}^2}$ ($16$ tiles, not extended) $[\mathrm{s}]$: & 73.35,0.47 & 39.92,0.35 & 29.24,0.37 & 24.65,0.42\\
        Time for $\smash{1 \, \mathrm{deg}^2 }$ ($16$ tiles, sides extended by $1.5$) $[\mathrm{s}]$:  & 160.9,3.9 & 83.71,0.49 & 57.40,0.48 & 41.16,0.35\\

        \bottomrule
        \end{tabular*}

\end{table*}

\begin{figure}
\centerline{\includegraphics[width=\columnwidth, trim= 0mm 0mm 0mm 0mm,clip]{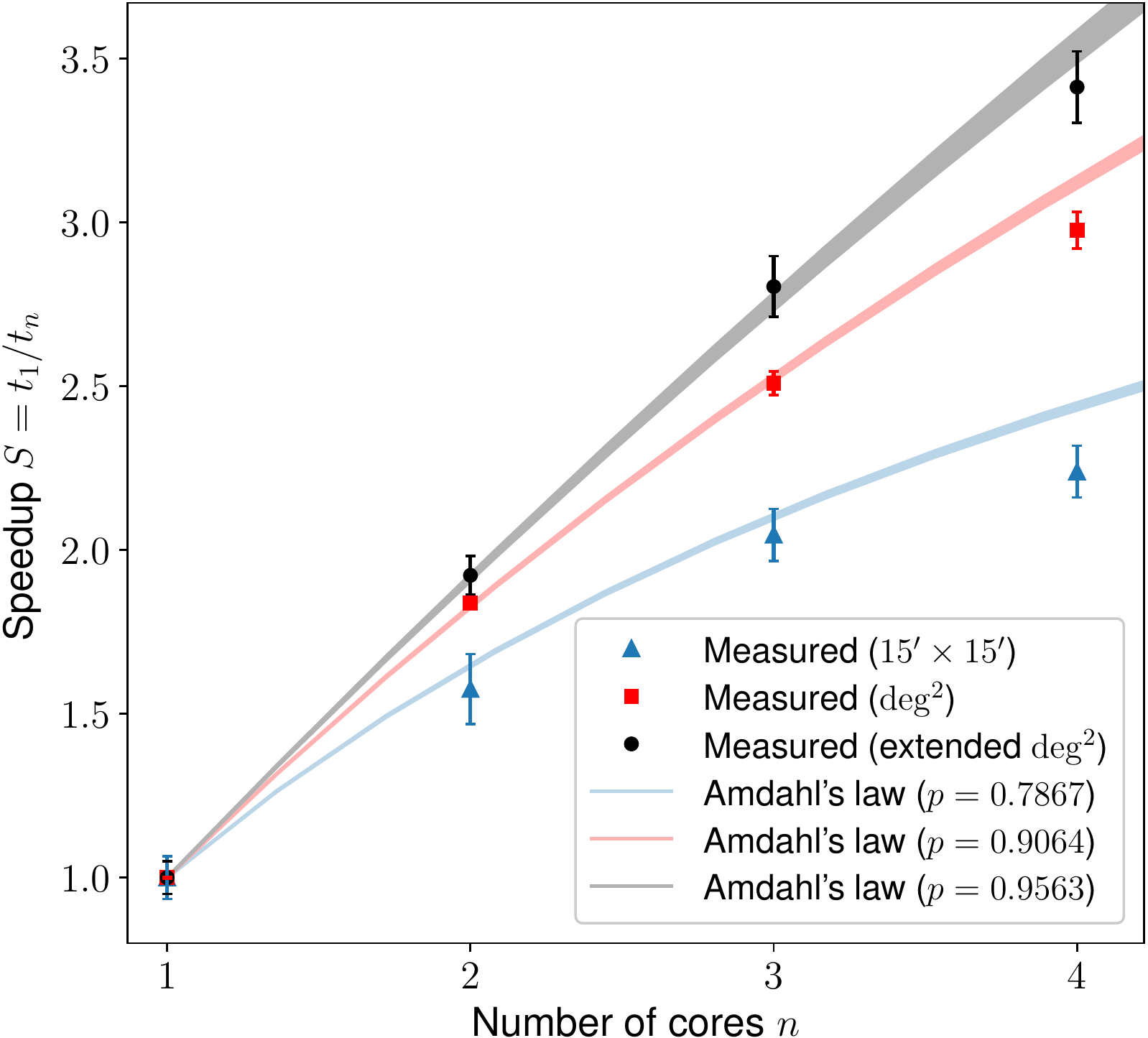} \hfill}
\caption{\label{fig:amdahl}Measured versus ideal speed-up of the computation time for the application mode as a function of CPU cores. The expected ideal speed-up is derived from Amdahl's law for parallel portions $p$ determined by direct time measurements.}
\end{figure}

For the application mode, we measured the total process runtime averaged over 10 runs. The results are shown in Table~\ref{table:perf2}. Here, the parallel portion is determined primarily by the search area and not expected to change significantly when applying different settings. Thus, we performed milisecond-precision measurements for two different regions -- a 'local cutout' region of $15' \times 15'$, without temporary buffer, and a $1\,\mathrm{deg^2}$ region, the latter both with and without temporary extension.
Since \emph{EasyCritics} spends most of its runtime in application mode, we perform a more detailed analysis of the gains that can be expected from the use of parallelization and how this behavior scales with the problem size. In particular, we evaluate and compare the speed-up for $n$ cores with respect to the single-core performance not only for both regions, but also against the theoretically expected ideal\footnote{If ignoring a variety of additional factors, such as communication overheads, load imbalance or super-linear speed-up.} speed-up proposed by \cite{amdahl1967}:
\begin{align}
        S_n = \frac 1{1 - p + \frac pn}.
\end{align}
Here, the parallel fraction is set to values derived from direct measurements of the sequential and parallel workload, which are $p = 0.7867 \pm 0.0024$ for the $15'\times15'$ region, $p = 0.9064 \pm 0.0023$ for the $1 \,\mathrm{deg^2}$ region and $p = 0.9563 \pm 0.0052$ for the temporarily extended $1 \,\mathrm{deg^2}$ region.

The measured speed-ups are shown together with the ideal Amdahl speed-ups in Fig.~\ref{fig:amdahl}. The error bars presented label the statistical errors. Note that the systematic uncertainties due to factors such as the number or luminosity density of galaxies in the volume or due to numerical settings can be larger. The data suggest a good qualitative agreement with the expectation, although small systematic deviations can be observed, which are most likely due to parallel overheads.
As expected, the speed-up improves for a larger area due to the increased parallel portion, indicating a partial weak performance scalability as well. The latter is of interest since \emph{EasyCritics} is intended for the application to very large datasets.

%%%%%%%%%%%%%%%%%%%%%%%%%%%%%%%%%%%%%%%%%%%%%%%%%%

% Don't change these lines
\bsp	% typesetting comment
\label{lastpage}
\end{document}